\documentclass[twocolumn,times,twocolappendix]{aastex631}
\usepackage{newtxmath}
\usepackage{amsmath}
\usepackage{enumerate}
\usepackage[english]{babel}
\addto\extrasenglish{

}

\DeclareRobustCommand{\VAN}[3]{#2}
\let\VANthebibliography\thebibliography
\def\thebibliography{\DeclareRobustCommand{\VAN}[3]{##3}\VANthebibliography}

\usepackage{tikz}
\usetikzlibrary{shapes.geometric, shapes.symbols, shapes.misc, arrows}

\tikzstyle{input} = [rounded rectangle, minimum width=3cm, minimum height=1cm,text centered, draw=black, fill=yellow!20]
\tikzstyle{process} = [ellipse, minimum width=3cm, minimum height=1cm, text centered, draw=black, fill=blue!20]
\tikzstyle{decision} = [signal, signal to=east and west, minimum width=3cm, minimum height=1cm, text centered, draw=black, fill=red!20]
\tikzstyle{output} = [rectangle, minimum width=3cm, minimum height=1cm,text centered, draw=black, fill=black!20]
\tikzstyle{end} = [rounded rectangle, minimum width=3cm, minimum height=1cm,text centered, draw=black, fill=green!20]
\tikzstyle{arrow} = [thick,->,>=stealth]

\tikzset{
  mylabel/.style = {font=\footnotesize, midway, fill=white, anchor=center}
}

\usepackage{calc}

\begin{document}

\title{The initial-to-final mass relation of white dwarfs in intermediate-separation binaries}


\author[0009-0002-9137-0631]{Oren Ironi}
\affiliation{Department of particle physics and astrophysics, Weizmann Institute of Science, Rehovot 7610001, Israel}
\correspondingauthor{Oren Ironi}
\email{oren.ironi@weizmann.ac.il}

\author[0000-0001-6760-3074]{Sagi Ben-Ami}
\affiliation{Department of particle physics and astrophysics, Weizmann Institute of Science, Rehovot 7610001, Israel}

\author[0000-0002-0430-7793]{Na'ama Hallakoun}
\affiliation{Department of particle physics and astrophysics, Weizmann Institute of Science, Rehovot 7610001, Israel}

\author[0000-0001-9298-8068]{Sahar Shahaf}
\affiliation{Department of particle physics and astrophysics, Weizmann Institute of Science, Rehovot 7610001, Israel}

\begin{abstract}
We examine the applicability of the initial-to-final mass relation (IFMR) for white dwarfs (WDs) in intermediate-separation binary systems ($\sim$1\,AU), using astrometric binaries identified in open clusters from Gaia DR3. A careful analysis of the astrometric orbits and spectral energy distributions isolates 33 main-sequence (MS) stars with highly likely WD companions. By combining cluster age estimates, dynamically measured WD masses, and, where available, WD cooling temperatures, we derive progenitor masses for 26 WD candidates. Our analysis suggests the presence of two distinct WD populations: (i) low-mass WDs, likely shaped by binary interactions during the progenitor's red-giant phase; and (ii) `spender' WDs, which experienced higher-than-expected mass loss and have progenitor masses above the IFMR predictions.  The rest of the candidates, referred to as the ‘others’, represent systems with inconclusive formation mechanisms. We suggest that at least some of these systems might be hierarchical triples, where the companion to the MS is a double WD or a double-WD merger product. However, follow-up studies are required to determine the nature of each case. These results highlight significant deviations from the IFMR derived for isolated WDs, emphasizing the role of binary evolution. Follow-up observations, particularly in the far-ultraviolet, are crucial for refining these findings and advancing our understanding of mass transfer processes and binary evolution pathways.
\end{abstract}

\section{Introduction} \label{sec:intro}
White dwarfs (WDs) represent the final evolutionary phase for the majority of stars ($\geq 97\%$; \citealt{althaus_97percent}). As such, the WD population plays a central role in numerous astrophysical processes. Still, despite their prevalence and importance, key aspects of WD formation and evolution processes are not fully understood (e.g., \citealt{WD_challenges_review}). One example is the relationship between a WD's mass and that of its progenitor. This dependence is known as the initial-to-final mass relation (IFMR).

The IFMR provides crucial insights into stellar evolution and is often used as a reference in population synthesis codes \citep[e.g.,][]{Cataln_2008, Toonen_2014}. Over the years, numerous studies have sought to estimate this relation, for example by using spectroscopic observations of WDs associated with open star clusters (e.g., \citealt{Cummings_2018}; but see other methods by \citealt{turnoff-wd, barnett_h_deficient, cunningham2024}). A common characteristic of most such studies is their reliance on either single WDs or members of binaries with orbits sufficiently wide to avoid significant interaction (\citealt{turnoff-wd}). While effective, these approaches do not account for the properties of WDs with binary companions close enough for interaction to affect their evolution significantly. This limitation is particularly relevant given the ubiquity of binary systems in stellar populations \citep[e.g.,][]{stellar_multiplicity}.

Over the past decade, extensive surveys have enabled significant progress in studying WDs in binary systems. Large datasets of binaries with well-characterized orbital properties have facilitated investigating systems that experienced interactions, bridging the gap between isolated WDs and those shaped by extreme processes such as common-envelope evolution (\citealt{ELM_survey,tight_binary_ifmr}). However, these surveys have primarily targeted close binaries, with separations on the order of a Solar radius, or wide binaries, with separations spanning hundreds of AU where interactions are negligible. The third data release from the Gaia mission recently revealed a previously unrecognized WD population with main-sequence (MS) companions at intermediate separations, around $1\,$AU. This discovery, within an unexplored region of parameter space, offers a rare opportunity to advance our understanding of binary formation and evolution processes \citep{Shahaf_2024,mass_deficit,post_CE_binaries,Rekhi_2024, li_WD_gaia}. We expect this intermediate separation population to experience less mass loss than the common-envelope systems but more mass loss relative to the wide binaries.

This study utilizes the MS+WD population discovered by Gaia to explore the role of binary interactions in WD formation. The analysis focuses on systems where the ages of the binary system can be constrained, enabling the estimation of WD progenitor masses. To this end, we construct and analyze a robust sample of MS+WD binaries in open clusters, where the orbital solution tightly determines the WD mass, and the cluster association establishes the system's age.

The paper is structured as follows: In \autoref{sec:sample}, we detail the catalogs and methods used to identify and compile the sample. \autoref{sec:analysis} details the analysis process used to constrain the properties of the secondary star in these binary systems, demonstrating, for instance, that the companion cannot be an MS star. Progenitor mass estimates for the final MS+WD candidates are presented in \autoref{sec:ifmr}. \autoref{sec:sample_properties} examines the orbital parameters of the candidates, focusing on the period-secondary star mass relationship, and the influence of selection biases on our sample. \autoref{sec:discussion} interprets the results, exploring evolutionary scenarios informed by the IFMR and addressing inconsistencies observed in some candidates. Finally, \autoref{sec:conclusions} summarizes the study and outlines potential directions for future research.

\section{Sample Selection}\label{sec:sample}

We construct our sample using two independent open-cluster catalogs. The \citet{Hunt_2023} catalog, based on Gaia DR3, identifies 7167 open clusters, of which we use 4105 that are flagged as highly reliable. The \citet{Dias_2021} catalog, based on Gaia DR2, provides homogeneous parameters for 1743 clusters derived through PAdova tRieste Stellar Evolutionary Code \citep[PARSEC;][]{parsec1,parsec2,parsec3,parsec4,parsec5,parsec6,parsec7} isochrone fitting. The clusters and individual members in the two lists are not mutually exclusive. We apply the same selection criteria for cluster members to both datasets: $\text{parallax}/\text{parallax error} > 10$ and $\text{astrometric fidelity} > 0.5$ \citep[see][]{Rybizki_fidelity}. 

We integrate data from the Gaia non-single star catalog \citep[NSS;][]{gaia_nss}, focusing on astrometric binaries with solution types \texttt{Orbital} or \texttt{AstroSpectroSB1}. We impose quality cuts on the orbital parameters, according to \citet{halbwachs2022gaia}. The sample is comprised of binaries with orbital periods of 100–1000 days, equivalent to intermediate orbital separations on the order of $\sim 1$\,AU. These binaries were then cross-matched with our list of open cluster members that have membership probability greater than $90\%$. 

The selection procedure yielded 283 binary systems distributed in 218 different open clusters. 
An outline of the process, along with subsequent steps of the analysis, is depicted in \autoref{fig:full_flowchart}. A more elaborated description of the sample selection procedure appears in \autoref{appendix:sample}.

\newsavebox{\flowchartbox}
\sbox{\flowchartbox}{
    \resizebox{\columnwidth}{!}{%
        \begin{tikzpicture}[node distance=2cm]
            
            \node (nss) [input] {NSS catalog};
            \node (hunt) [input, right of=nss, xshift=2cm, align=center] {Hunt \& Reffert\\ catalog};
            \node (dias) [input, right of=hunt, xshift=2cm, align=center] {Dias et al.\\ catalog};
            
            \node (reliable_orbits) [decision, below of=nss, align=center] {Astrometric binaries \\ Reliable orbits};
            \node (reliable_clusters) [decision, below of=hunt, align=center] {Reliable\\clusters};
            \node (cluster_criteria) [decision, below of=dias, align=center, yshift=-1.5cm] {Reliable photometry \\ and astrometry \\ Secure cluster association};
            
            \node (candidates) [output, below of=cluster_criteria, yshift=-0.5cm] {283 candidates};
            \node (cluster_analysis) [process, left of=candidates, xshift=-2cm, align=center] {Cluster analysis \\ Primary mass \\ AMRF triage};

            \node (discard_msms) [decision, below of=cluster_analysis, yshift=-0.5cm, align=center] {Discard MS+MS \\ $M_2 < 1.4$ M$_\odot$\\ Either $M_2/\delta M_2 > 20$\\ Or $M_1/\delta M_1 > 3\times q/\delta q$ };
            
            \node (mswd_candidates) [output, below of=discard_msms] {40 candidates};
            \node (sed_analysis) [process, below of=mswd_candidates] {SED analysis};
            \node (ir_excess)[decision, below of= sed_analysis]{IR excess?};
            \node (hierarchical_triple)[end, left of= ir_excess, xshift=-2cm]{Hierarchical triple MS};
            \node (mswd_candidates2)[output, below of= ir_excess]{34 MS+WD candidates};
            \node (uv_data) [decision, below of=mswd_candidates2] {UV data exist?};
            \node (uv_excess) [decision, below of=uv_data, xshift=2cm, yshift=-0.5cm] {UV excess?};
            \node (ms+sdb)[end, above right of=uv_excess, xshift=1.0cm,yshift=0.5cm]{1 MS+SD};
            \node (confirmed_mswd) [output, below of=uv_excess, xshift=2cm, align=center, yshift=-0.5cm] {8 confirmed MS+WDs \\ $T_\text{WD}$ known};
            \node (mswd_noexcess) [output, below of=uv_excess, align=center, yshift=-0.5cm, xshift=-2cm] {9 candidate MS+WDs\\ no UV excess};
            \node (mswd_nouv) [output, below of=uv_data, align=center, xshift=-3cm, yshift=-0.5cm] {16 candidate MS+WDs \\ no UV data};
            \node (progenitor_mass) [process, below of=confirmed_mswd, xshift=2cm ,align=center] {4 progenitor masses \\ $\tau_\text{MS} = \tau_\text{tot} - \tau_\text{cool}$};
            \node (kroupa_prior) [process, below of=mswd_noexcess, xshift=-2cm, align=center] {22 progenitor masses \\ IMF prior};
            \node (excluded)[end, left of=progenitor_mass,xshift=-2cm]{7 excluded};
            \node (ifmr) [end, below of=progenitor_mass, xshift=-3cm] {IFMR};
            
            \draw [arrow] (nss) -- (reliable_orbits);
            \draw [arrow] (hunt) -- (reliable_clusters);
            \draw [arrow] (dias) -- (cluster_criteria);
            \draw [arrow] (reliable_orbits) |- (cluster_criteria);
            \draw [arrow] (reliable_clusters) |- (cluster_criteria);
            
            \draw [arrow] (cluster_criteria) -- (candidates);
            \draw [arrow] (candidates) -- (cluster_analysis);
            \draw [arrow] (cluster_analysis) -- (discard_msms);
            
            \draw [arrow] (discard_msms) -- (mswd_candidates);
            \draw [arrow] (mswd_candidates) -- (sed_analysis);
            \draw [arrow] (sed_analysis) -- (ir_excess);
            \draw [arrow] (ir_excess) -- (mswd_candidates2) node[mylabel] {No};
            \draw [arrow] (ir_excess) -- (hierarchical_triple) node[mylabel] {Yes};
            \draw [arrow] (mswd_candidates2) -- (uv_data);
            \draw [arrow] (uv_data) -- node [mylabel] {No} (mswd_nouv);
            \draw [arrow] (uv_data) -- node [mylabel] {Yes} (uv_excess);
            \draw [arrow] (uv_excess) -- node [mylabel] {Yes} (confirmed_mswd);
            \draw [arrow] (uv_excess) -- node[mylabel]{Yes} (ms+sdb);
            \draw [arrow] (uv_excess) -- node [mylabel] {No} (mswd_noexcess);
            \draw [arrow] (confirmed_mswd) -- (progenitor_mass);
            \draw [arrow] (mswd_noexcess) -- (excluded);
            \draw [arrow] (confirmed_mswd) -- (excluded);
            \draw [arrow] (mswd_noexcess) -- (kroupa_prior);
            \draw [arrow] (mswd_nouv) -- (kroupa_prior);
            \draw [arrow] (progenitor_mass) -- (ifmr);
            \draw [arrow] (kroupa_prior) -- (ifmr);
    
        \end{tikzpicture}
        }
}

\newsavebox{\legendbox}
\sbox{\legendbox}{\fbox{%
    \resizebox{0.3\columnwidth}{!}{%
        \begin{tikzpicture}[node distance=0.7cm]
            \node (inputs) [input, minimum width=1.5cm, minimum height=0.5cm]{};
            \node (inputs_txt) [right of=inputs, xshift=1.5cm]{Inputs};
            \node (cuts) [decision, minimum width=1.5cm, minimum height=0.5cm, below of=inputs] {};
            \node (cuts_txt) [below of=inputs_txt] {Cuts};
            
            \node (steps) [process, minimum width=1.5cm, minimum height=0.5cm, below of=cuts] {};
            \node (steps_txt) [below of=cuts_txt] {Analysis};
            
            \node (samples) [output, minimum width=1.5cm, minimum height=0.5cm, below of=steps] {};
            \node (samples_txt) [below of=steps_txt] {Sample size};
            \node (endpoints) [end, below of= samples,minimum width=1.5cm, minimum height=0.5cm]{};
            \node[below of=samples_txt]{End point};
        \end{tikzpicture}}
        }
    }

\newlength{\flowchartboxwidth}
\setlength{\flowchartboxwidth}{\wd\flowchartbox}
\newlength{\flowchartboxheight}
\setlength{\flowchartboxheight}{\ht\flowchartbox}

\newlength{\legendboxwidth}
\setlength{\legendboxwidth}{\wd\legendbox}
\newlength{\legendboxheight}
\setlength{\legendboxheight}{\ht\legendbox}

\newlength{\posx}
\setlength{\posx}{0.5\flowchartboxwidth - 0.5\legendboxwidth}
        
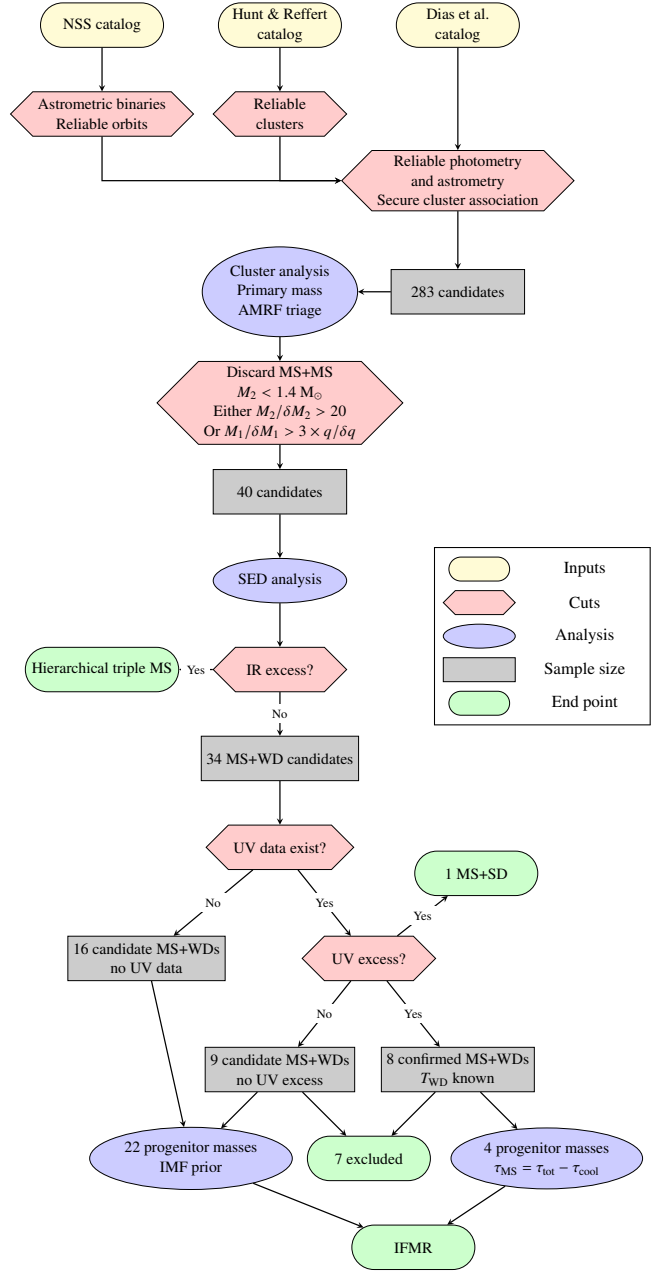
\begin{figure}
    \centering
    \begin{tikzpicture}
        \node (flowchart) at (0,0) {\usebox{\flowchartbox}};
        \node at (\posx,0) [align=right] {\usebox{\legendbox}};
    \end{tikzpicture}
    \caption{Flow chart describing the entire work from initial target selection to the final IFMR. Cuts are shown in pink hexagons, steps of analysis in blue ellipses, sample size during different steps in grey rectangles (see legend).}
    \label{fig:full_flowchart}
\end{figure}

\section{The nature of the secondary star} \label{sec:analysis}

The sample comprises 283 binaries with reliable orbital solutions and high probability cluster membership. In the following section, we describe the process used to exclude systems in which the companion is consistent with being luminous, for example an MS star. The final sample consists of $34$ systems where the companion is likely a compact object. Given the rarity of other compact stellar objects, these companions are most plausibly WDs, and we thus classify these $34$ systems as MS+WD candidates.

\subsection{Cluster Parameters} \label{subsec:cluster_analysis}
Accurate estimates of the system's age, extinction, and metallicity are essential for modeling the spectral energy distribution (SED) and mass-to-light ratio of the binary components. These estimates are crucial for constraining the nature of the faint secondary companion.

While such estimates are challenging for field systems, they are more accessible for our sample. The color-magnitude diagram (CMD) of open clusters provides a reliable foundation for determining these parameters:
We fit the observed CMDs of all 218 clusters with PARSEC v1.2S isochrones,\footnote{Available online: \url{http://stev.oapd.inaf.it/cgi-bin/cmd}} obtaining estimates for the age, extinction, and metallicity of each system. This procedure employs the code developed by \citet{Dias_fitting_code}.\footnote{Available online: \url{https://github.com/hektor-monteiro/OCFit}} The same code was also used by \citet{Dias_2021} in compiling their catalog, and we have updated it to utilize Gaia DR3 data.

There are $30$ clusters in our sample that appear both in the \citet{Dias_2021} and \citet{Hunt_2023} catalogs. Due to differences in membership lists, we fit these duplicate clusters twice using the weighted averages of the parameters as the best fit. For most cases, the cluster parameters we derive from both catalogs agree within 2 sigma. However, in $4$ cases, the analysis yielded conflicted results; these clusters were therefore removed from the sample. For a detailed description of the fitting algorithm, along with a comparison plot of \citet{Hunt_2023} cluster ages versus \citet{Dias_2021} cluster ages, refer to \autoref{appendix:isochrones} and \autoref{fig:cluster_comparison}.

\subsection{Primary Mass Estimate} \label{subsec:mass_analysis}
To constrain the properties of the companion, we require that the primary be an MS star with a well-determined mass. The primary mass, $M_1$, is derived assuming that the primary dominates the system's flux in the Gaia passbands. This assumption holds for MS+WD binaries, as shown by representative SED models, which demonstrate that the WD's flux contribution in these passbands is negligible compared to that of the MS primary (see \autoref{subsec:sed_analysis}). Furthermore, \citet{Yamaguchi2024} independently supports this assumption for a similar sample. Their joint analysis of Gaia astrometry and RV follow-up constrains the flux ratio to an upper limit of $2\%$ in Gaia's $G$ band.

However, in the case of binaries of two MS stars, this assumption does not hold and may result in biased estimates of $M_1$. As discussed in \autoref{subsec:amrf_analysis}, the resulting discrepancies help filter out these systems. We use the best-fitting PARSEC isochrone, the cluster extinction, and the observed Gaia color and magnitude, $G_\text{BP}-G_\text{RP}$, $G$, to interpolate the photometric mass of each source, and run a Monte Carlo simulation to estimate the uncertainty $\delta M_1$. See \autoref{appendix:photometric_mass} for a step-by-step outline of this procedure.

\subsection{Astrometric Triage} \label{subsec:amrf_analysis}

The astrometric triage method \citep{Shahaf_2019} uses the observed photocentric orbit of unresolved binary systems to identify those with compact secondaries. If the companion is luminous, the photocenter shifts from the position of the primary toward the center of mass, causing the observed orbit to appear smaller than its true size. Assuming the system consists of two MS stars, two extreme cases can be considered: (i) the secondary mass is negligible compared to the primary, and (ii) the secondary mass is roughly equal to that of the primary. In both cases, the angular semi-major axis of the photocentric orbit goes to zero. This implies a maximal value, corresponding to an intermediate mass ratio between zero and one, where the photocentric orbit reaches its largest size. If the observed orbit exceeds this limit, the companion cannot be a luminous MS star.

The triage technique relies on the \emph{astrometric mass ratio function} (AMRF), defined as
\begin{equation} \label{eq:amrf_obs}
    \mathcal{A} = \frac{\alpha_0}{\varpi}\left(\frac{M_1}{M_\odot}\right)^{-1/3}\left(\frac{P}{\text{yr}}\right)^{-2/3},
\end{equation}
where $P$ is the orbital period, $\varpi$ is the parallax, and $\alpha_0$ is the angular semi-major axis of the photocentric orbit. These parameters are derived from the Thiele-Innes coefficients provided in the NSS catalog \citep{halbwachs2022gaia}. 
The AMRF is linked to the system's mass and flux ratios through
\begin{equation} \label{eq:amrf_variables}
    \mathcal{A} = \frac{q}{(1+q)^{2/3}}\left(1 - \frac{\mathcal{S}(1+q)}{q(1+\mathcal{S})}\right),
\end{equation}
where $q = M_2/M_1$ is the mass ratio, and $\mathcal{S} = F_2/F_1$ is the flux ratio. Using the estimated age, mass, and metallicity of the primary, we model $\mathcal{S}$ as a function of $q$ and calculate the maximal value for an MS companion, $\mathcal{A}_{\textit{MS}}$. 

If the observed AMRF is larger than $\mathcal{A}_{\textit{MS}}$ we rule out the possibility of a single MS companion. This is done via a Monte Carlo simulation to account for measurement uncertainties: In each iteration, we sample the observed AMRF using the uncertainties in the orbital parameters and compute the limiting value, incorporating uncertainties in age, metallicity, and extinction. The fraction of iterations where $\mathcal{A}$ falls below $\mathcal{A}_{\textit{MS}}$ defines the system's \texttt{class-I probability} \citep{Shahaf_2023}. We retain 53 systems with a \texttt{class-I probability} below $10\%$. After rejecting the possibility that the companion is a single MS star, the most plausible remaining possibilities are that its faint companion is a compact object or a close binary of two low-mass MS stars. The technical details required for the reproducibility of this selection process are provided in \autoref{appendix:amrf}.

This triage approach has been successfully applied to Gaia binaries to identify systems with compact companions, as shown in \citet{Shahaf_2023}, \citet{Shahaf_2024}, and \citet{nss_amrf_official}. Ground based follow-up campaigns validated the orbit and triage classification (\citealt{Yamaguchi2024}). This work, as opposed to the above mentioned studies in which $\mathcal{A}_{\textit{MS}}$ was estimated globally across a range of ages and metallicities, calculates the limiting value individually for each system. This calculation accounts for $M_1$, [$\mathrm{Fe/H}$], and $A_V$, and is compared directly to the observed AMRF value.

Assuming that the companion is not a single MS star, we now estimate the mass of the secondary under the assumption that its contribution to the total flux is negligible. We set $\mathcal{S}$ to $0$ and solve \autoref{eq:amrf_variables} to determine the mass ratio, $q$. We exclude three systems where the companion mass exceeds the Chandrasekhar limit and discard systems with large relative errors in $M_2$, by demanding either $M_1 / \delta M_1 > 3 \times q / \delta q$ or $M_2 / \delta M_2 > 20$. After applying these criteria, 40 systems remain in the sample. In the following subsection, the assumption that the secondary flux is negligible is validated, and if needed, the secondary mass is estimated again (see \autoref{subsec:sed_analysis}).

To summarize, this step in our analysis yielded a set of 40 binaries in which the secondary companion is likely not a single MS star. The most plausible possibilities for its nature are that it is a compact object or a close binary of two MS stars, forming a hierarchical triple system.

\subsection{SED Analysis} \label{subsec:sed_analysis}

We use spectral energy distribution (SED) to further constrain the companion's nature. Analyzing the SED of the 40 binaries from \autoref{subsec:amrf_analysis}, we estimate the dominant contribution of the MS primary and test for excess emission in either the infrared (IR) or ultraviolet (UV) bands. IR excess indicates the presence of low-mass MS stars, allowing us to identify and exclude 6 possible hierarchical triple MS systems from the sample. In other cases, residual flux in the UV reveals the contribution of the WD companion, providing a direct confirmation of its existence and constraining its effective temperature, $T_{\text{eff},2}$. This temperature, in turn, produces the WD cooling age. This subsection qualitatively outlines our analysis. See \autoref{appendix:sed} for technical details.

We construct the SED using multiple databases. When possible, we include GALEX far- and near-UV photometry ($FUV$ and $NUV$; \citealt{galex-mast,galex7}), which is available for 18 of the 40 systems in our sample and reported in \autoref{tab:candidate_table}. In the optical range, we use Gaia $G$, $G_{\text{BP}}$, and $G_{\text{RP}}$, along with Gaia synthetic photometry in Johnson-Kron-Cousins $UBVRI$ bands \citep{synthetic_photometry}. For the IR, we incorporate 2MASS $J$, $H$, $K_s$ \citep{2mass_catalog} and WISE $W1$, $W2$, $W3$ \citep{wise_catalog}. To ensure a homogeneous treatment of the entire sample, we avoided surveys with partial coverage (e.g., Pan-STARRS, SDSS). Although the reported photometric uncertainties, particularly in the optical, are small with signal-to-noise ratios often in the hundreds or thousands, stellar SED models are probably not precise to this level. Therefore, we impose a minimum uncertainty of $10\%$ for all photometric measurements to account for modeling limitations. 

\begin{figure}
    \centering
    \includegraphics[width=\columnwidth]{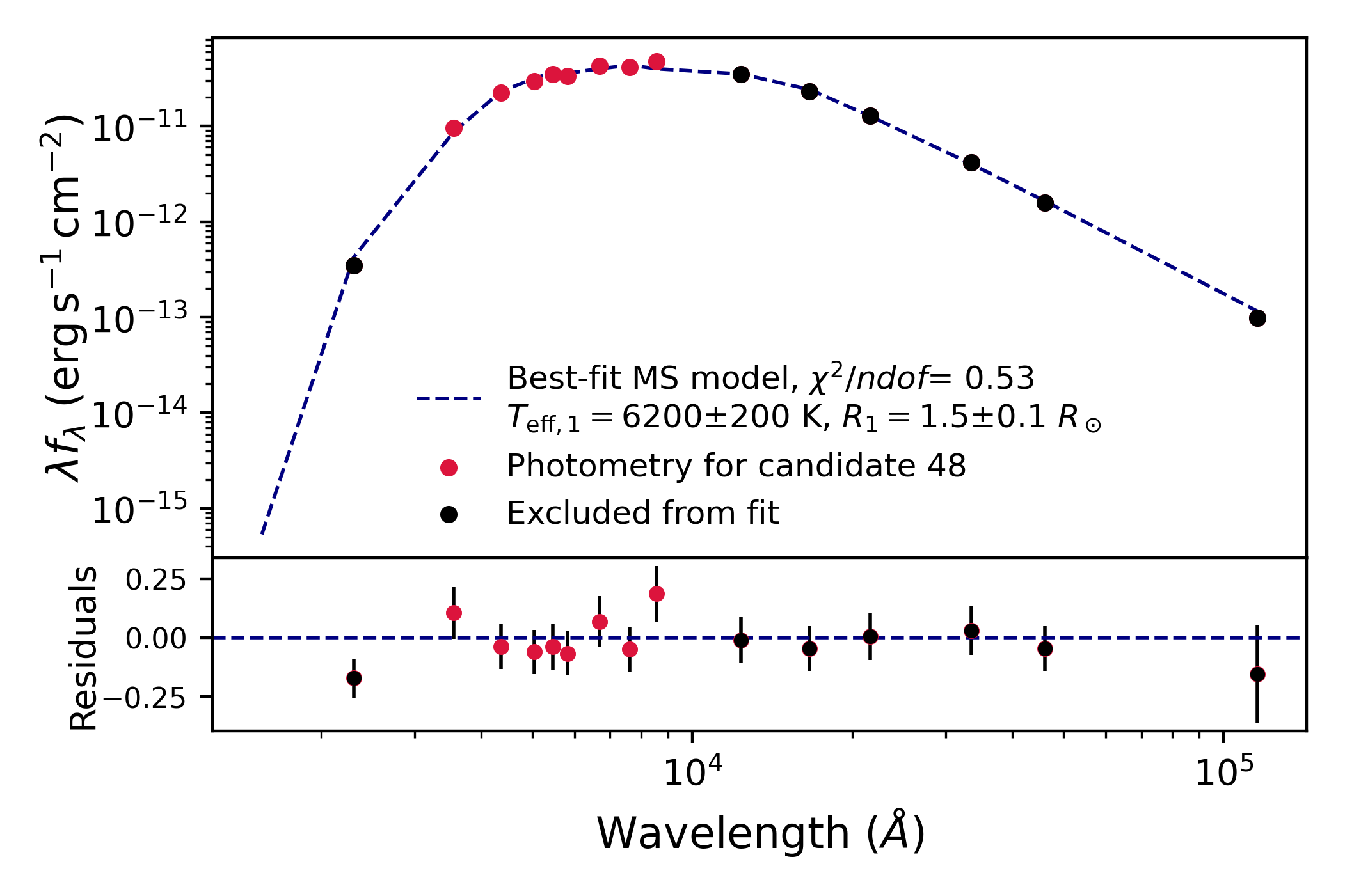}
    \caption{Fit of optical SED to a single MS star. The black data points (GALEX $NUV$, 2MASS $J,H,Ks$, WISE $W1,W2,W3$) were excluded from the fit. The dashed line shows the synthetic photometry of the best-fitting MS model. 
    No IR or UV excess is seen in this fit. The residuals in the bottom panel are defined as $\left(f_\text{obs}-f_\text{model}\right)/f_\text{model}$.}
    \label{fig:kurucz_fit_no_excess}
\end{figure}

\begin{figure}
    \centering
    \includegraphics[width=\columnwidth]{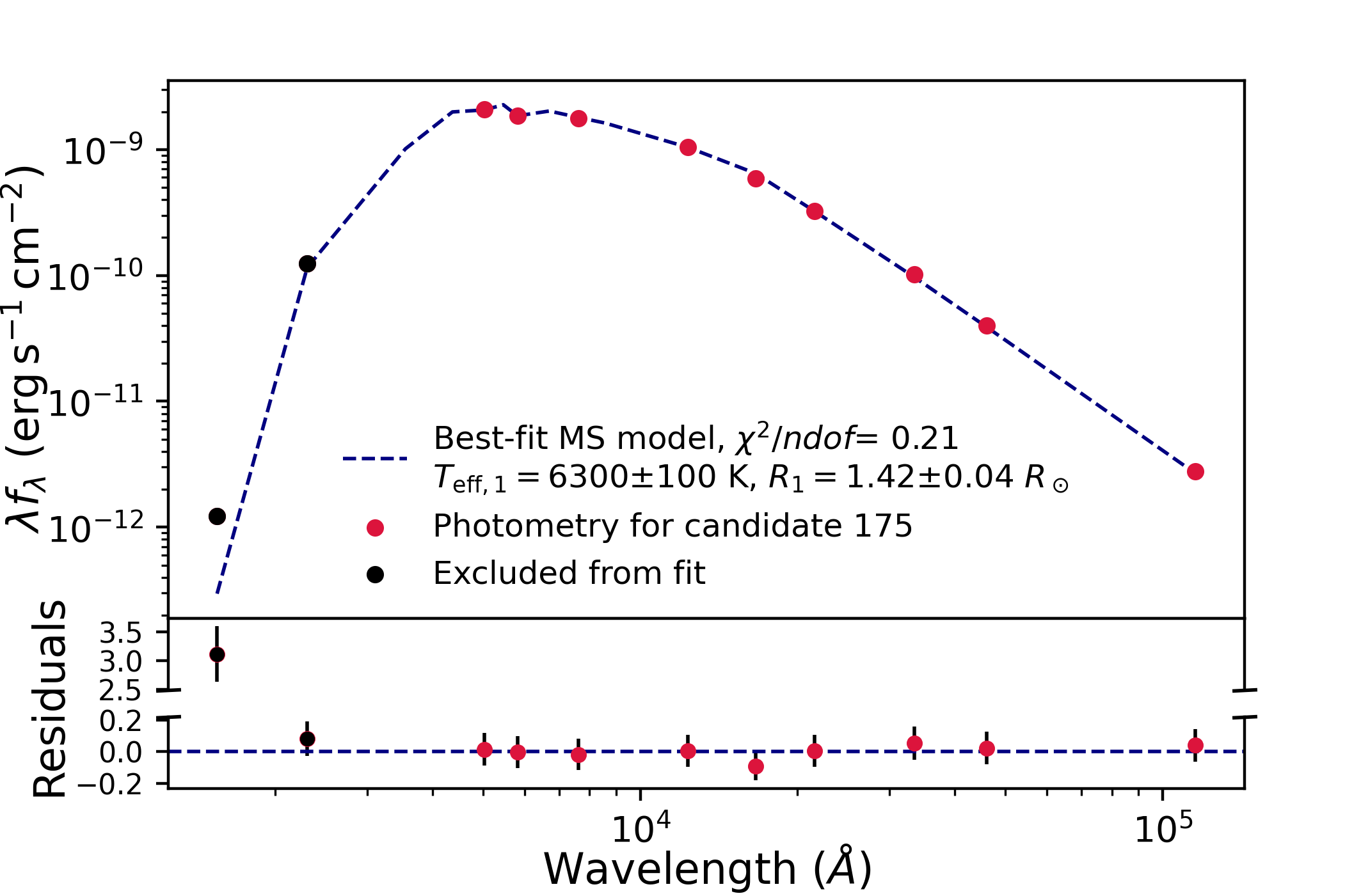}
    \caption{Fit of optical+IR SED to a single MS star. The black data points (GALEX $FUV,\,NUV$) were excluded from the fit. The dashed line shows the synthetic photometry of the best-fitting MS model. This target shows excess in the GALEX $FUV$ band. The residuals are defined as in \autoref{fig:kurucz_fit_no_excess}.}
    \label{fig:kurucz_fit_with_excess}
\end{figure}

We first fit the SED with a single MS star model, using synthetic photometry based on the atmospheric models of \citet{Kurucz_Castelli}.\footnote{The \citet{Kurucz_Castelli} models can be downloaded from the Spanish Virtual Observatory website, under the `synthetic photometry' tab: \url{http://svo2.cab.inta-csic.es/theory/newov2/syph.php}.} In this initial step, we use only the optical bands where contamination by the companion is expected to be negligible. The single-star best-fitting model is our baseline for detecting UV or IR excess (see \autoref{fig:kurucz_fit_no_excess}). In addition, the best-fit model estimates the primary star's radius and effective temperature. The single-object fit shows IR excess for 6 candidates, all lacking UV data. We repeat the fitting process for the remaining 34 systems, now also including the uncontaminated IR data, to refine our estimates for the primary's temperature and radius (see \autoref{fig:kurucz_fit_with_excess}). The SED fitting results are given in \autoref{tab:sed_results} in \autoref{appendix:sed}. 

\begin{figure}
    \centering
    \includegraphics[width=\columnwidth]{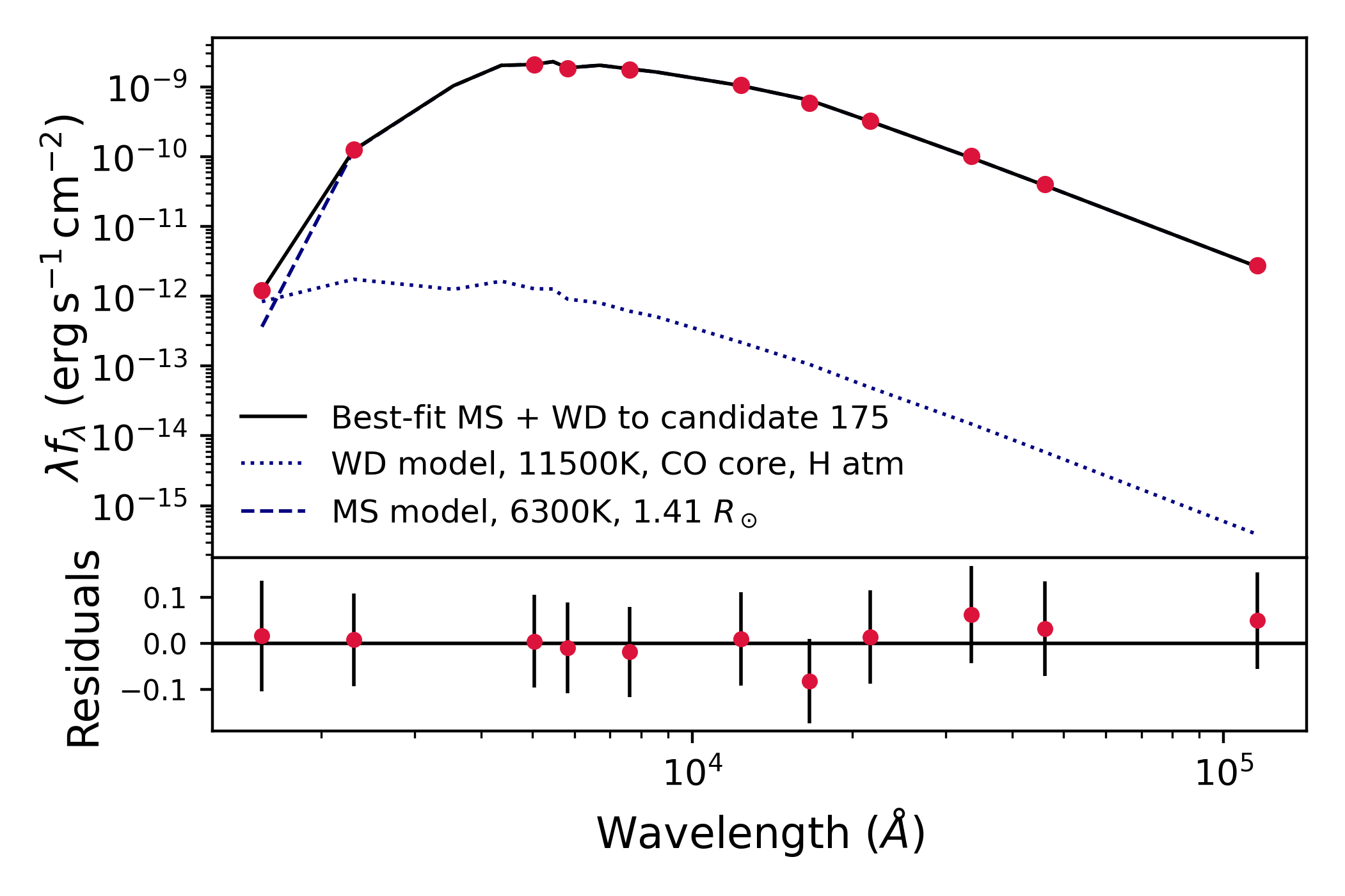}
    \caption{Best fit of the SED of candidate 175 to a combined MS+WD model. The data points from left to right are observed flux in $FUV\,,NUV,\,G_\text{BP},\,G,\,G_\text{RP},\,J,\,H,\,Ks,\,W1,\,W2,\,W3.$ The dashed, dotted, and solid lines show the synthetic photometry of the best-fitting MS, WD, and combined models. The residuals are defined as in \autoref{fig:kurucz_fit_no_excess}.}
    \label{fig:full_binary_fit}
\end{figure}

Of the 34 systems with no IR excess, 18 have available UV data (either $NUV$ or $FUV$; see \autoref{tab:candidate_table}). Out of these, $9$ systems show UV emission higher than expected from what is estimated based on our optical+IR fit: systems number 53, 133, 174, 175, 180, 236, 249, 281 and 283.
We fit the full SED of these systems with a combined MS+WD model using the \textsc{emcee} package \citep{emcee}. The fitted models are comprised of the MS synthetic photometry of \citet{Kurucz_Castelli} combined with the WD synthetic photometry by \citet{montreal_cooling_models}.\footnote{The models can be downloaded from \url{https://www.astro.umontreal.ca/~bergeron/CoolingModels/}.} We use WD cooling models and mass-radius relations, listed in \autoref{tab:massradius}, accounting for the expected WD core composition in different mass regimes. An example of a combined fit is shown in \autoref{fig:full_binary_fit}. For 8 candidates, the joint fit supports our claim for the existence of a WD companion and yields estimates for $T_{\text{eff},2}$. The remaining candidate, number 180, has a $NUV$ excess that is too large to be explained by a WD. It is excluded from the sample since our analysis suggests that the faint companion in this system is probably a hot subdwarf (SD); see \autoref{appendix:excluded_systems}.

The absence of UV excess in the remaining 9 systems cannot definitively rule out the presence of a WD. A sufficiently hot primary star could outshine a WD in this band, or the WD could be too cold to produce detectable excess. These considerations are specifically relevant for these 9 binaries, since only the $NUV$ band is available for them. 
In these cases, we set an upper limit on the WD temperature, $T_{\text{hottest}}$. This limit is defined such that a WD's expected $NUV$ flux equals $10\%$ of the observed flux in that band, assuming a hotter companion would produce a detectable $NUV$ excess. We use these upper limits and discuss them further in \autoref{subsec:ifmr_no_excess}.

\begin{deluxetable}{ccc}
\tablecaption{WD mass-radius relations and cooling tracks used in this work. \label{tab:massradius}}
\tablehead{\colhead{Mass range} & \colhead{Model} & \colhead{Reference}}
\startdata
 $ <0.45 M_\odot$ & He core & \citet{He_models} \\ 
 $0.45M_\odot - 1.1 M_\odot$ & CO core& \citet{montreal_cooling_models} \\  
 $1.1M_\odot- 1.4M_\odot$ & ONe core & \citet{ONe_models}   \enddata
\end{deluxetable}

\section{The Initial-to-Final Mass Relation} \label{sec:ifmr}
Thus far, using the astrometric triage and SED analysis, we identified 33 MS+WD candidates. Of these, 8 systems exhibit UV excess, 9 have UV data but no detectable excess, and 16 lack UV data. This section separately considers each situation, starting with the UV-excess candidates.

If UV data are available and excess emission is detected, we can estimate the temperature of the WD, and through it estimate how much time had passed since its formation. This procedure enables the inference of the mass of its progenitor star. We follow the methodology of \citet{Cummings_2018}, outlined in the following subsection. If UV excess is not detected (or UV data are not available), we cannot estimate the cooling age. In these cases, we constrain the mass of the progenitor using statistical considerations about the assumed properties of the population, assuming progenitor masses were drawn from the \citet{kroupa_imf} initial mass function (IMF), and using the upper limits on the WD temperature whenever possible.

Below we outline the process we used to derive, or constrain, the progenitor masses. The Gaia source IDs and the derived initial and final masses are provided in \autoref{tab:candidate_table}, that appears with the complete technical details in \autoref{appendix:ifmr}.

\subsection{UV Excess} \label{subsec:ifmr_uv_excess}

\begin{figure}
    \includegraphics[width=\columnwidth]{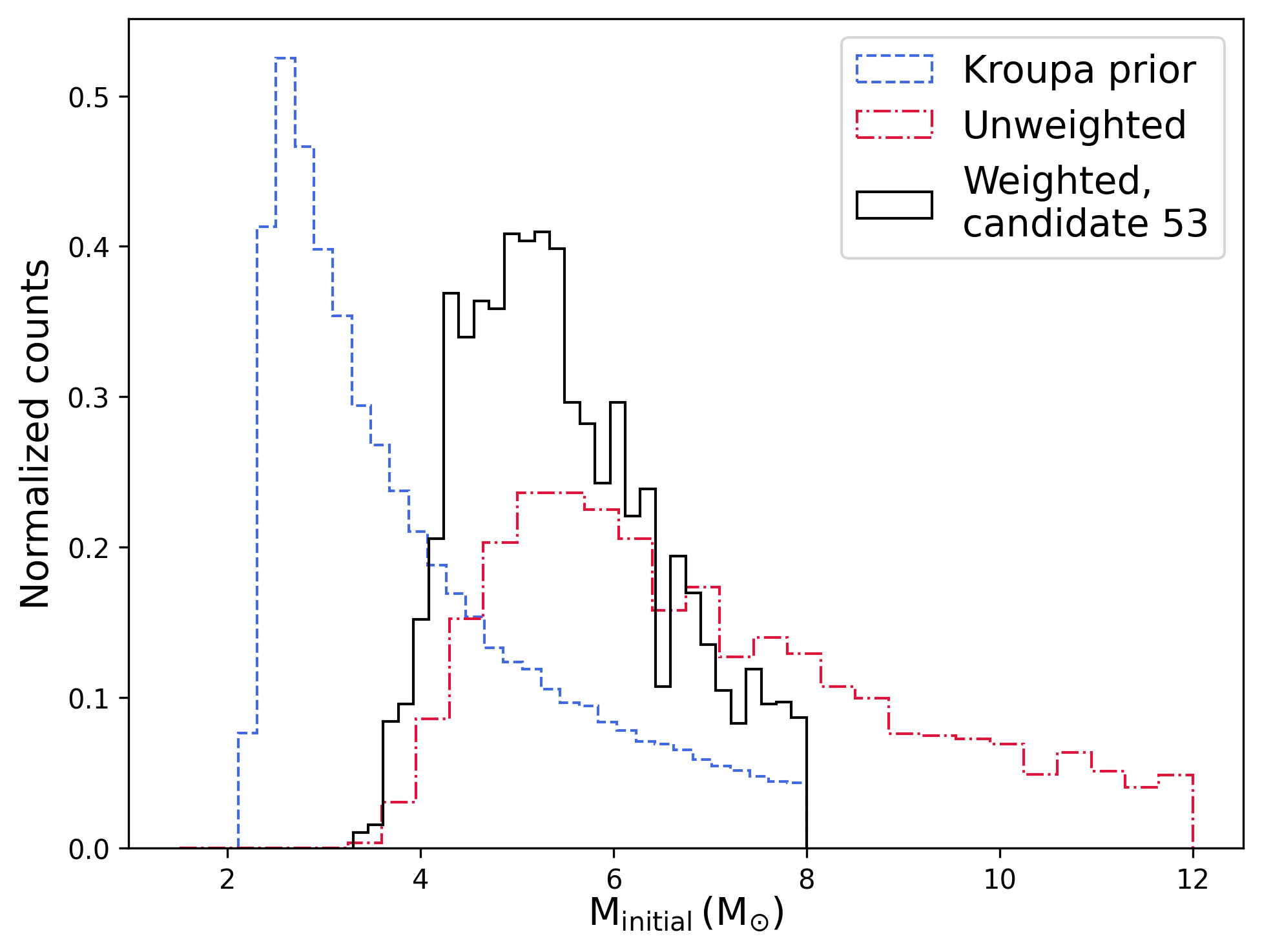}
    \caption{Final mass distribution for the progenitor of candidate 53. In dashed blue, we see the initial prior based on the IMF, in dot-dashed red the unweighted distribution that follows from the WD cooling temperature and the cluster age, and in solid black the latter distribution after applying weights based on the prior.}
    \label{fig:m_initial_posterior}
\end{figure}

The binary fit described in \autoref{subsec:sed_analysis} provided the posterior distribution of the effective temperature of the secondary star in eight confirmed MS+WD binaries. For these eight, we can provide a robust estimate of the progenitor mass using the posterior of $T_{\text{eff},2}$ and following the \citet{Cummings_2018} methodology:
\begin{enumerate}[i]
    \item Estimate the WD effective temperature using a WD cooling track to derive its cooling age, $\tau_\text{cool}$.
    \item Subtract the WD cooling age from the cluster age to get the progenitor lifetime, $\tau_\text{life} = \tau_\text{tot} - \tau_\text{cool}$.
    \item Use $\tau_{\text{life}}$ to derive the progenitor's mass, M$_\text{initial}$, using a PARSEC evolutionary track.\footnote{Similar to \citet{Cummings_2018}, we define the lifetime as the age of the star at the start of the asymptotic giant branch (AGB) phase. For each system, we use a track with the metallicity derived from its cluster.}
\end{enumerate}
We account for the measurement uncertainties, producing a distribution of $M_{\text{initial}}$ values, as the red histogram in \autoref{fig:m_initial_posterior} demonstrates.

The initial mass derivation process described above does not account for the expected occurrence of WD progenitors in the host cluster. To address this, we require the progenitor mass to satisfy $M_{\text{turnoff}} < M_{\text{initial}} < 8M_\odot$, with a probability density following the \citet{kroupa_imf} IMF with power index of $2.3$. These constraints are incorporated using a rejection sampling approach:
We construct a prior by sampling the IMF, generating 5000 values within the allowed range while accounting for uncertainties in the cluster age (see the blue histogram in \autoref{fig:m_initial_posterior}). Cooling-age-based estimates are then weighted by this prior to produce the posterior distribution of the initial mass, shown in black in \autoref{fig:m_initial_posterior}. The estimated value of the initial mass is taken as the median of the weighted samples, and the 16\textsuperscript{th} and 84\textsuperscript{th} percentiles are used as the corresponding confidence interval.

We were unable to derive the progenitor mass for 4 of these systems. In 3 cases, the derived cooling ages exceed the cluster age, even when accounting for random errors (candidates 174, 281, 283; see \autoref{appendix:excluded_systems}). Another system (candidate 133) is excluded because $M_\text{turnoff} > 8\,\text{M}_\odot$ and requires specific consideration. Further investigation suggests that false cluster associations and underestimated cluster ages may explain these inconsistencies.

\begin{figure*}
    \centering
    \includegraphics[width=0.7\textwidth]{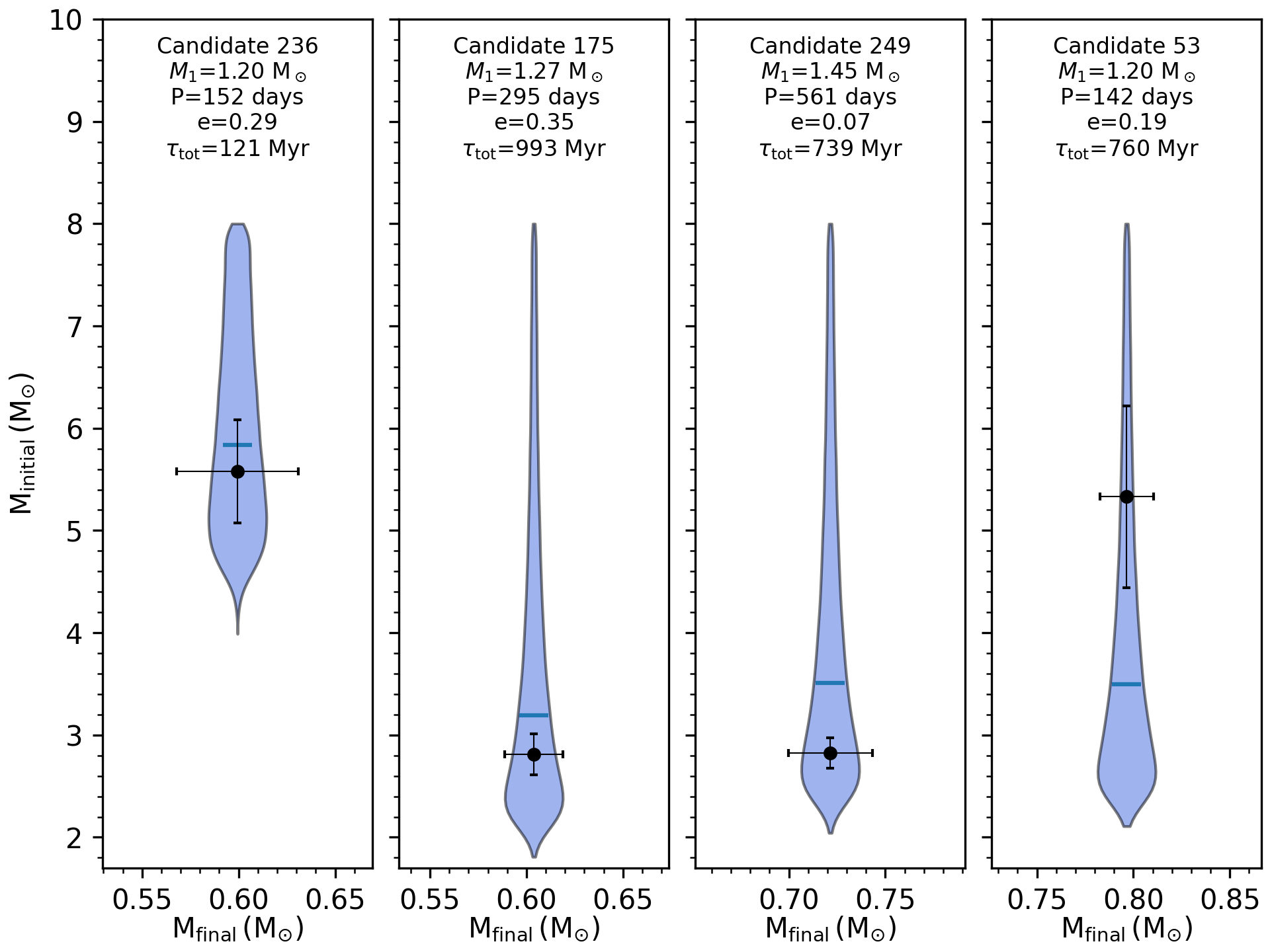}
    \caption{The initial versus final mass of four systems with UV excess and a successful binary fit (black error bars). The Kroupa-based prior (see text) is presented as a blue violin plot, with the median of the distribution indicated by a blue horizontal dash. Additional UV data are crucial for narrowing down the uncertainty on $M_\text{initial}$, evident from the small error bars for candidates 175 and 249 ($NUV+FUV$) relative to 53 ($FUV$ only) and 236 ($NUV$ only).}
    \label{fig:ifmr_uv_excess}
\end{figure*}

The remaining 4 systems show well-constrained fits to a joint MS+WD SED model, with UV emission attributed to the WD. The progenitor and WD remnant masses derived from these systems are presented in \autoref{fig:ifmr_uv_excess}, alongside the distribution obtained using only the IMF prior, without UV excess constraints. This comparison supports the use of an IMF-based prior for progenitor masses. Accordingly, in cases where UV excess is undetected or unavailable, the IMF prior is adopted.

\subsection{No UV Excess} \label{subsec:ifmr_no_excess}
For the 9 systems without UV excess, we assume the WD is outshined by its primary companion in the $NUV$ band. Using the IMF prior and upper limits on the WD temperature, we constrain their progenitor masses. The process is similar to the analysis of systems with detected UV excess but applies less restrictive constraints on the prior. As described in \autoref{subsec:sed_analysis}, $T_{\text{hottest}}$ sets a lower bound on $\tau_\text{cool}$, which, for a given $\tau_\text{tot}$, corresponds to a lower bound on $M_\text{initial}$. This bound is used as a lower cutoff to the IMF, analogous to the approach applied above with $M_\text{turnoff}$. Follow-up $FUV$ observations will be important to refine and validate these results.

For 6 targets, we obtained refined estimates of $M_\text{initial}$. However, a significant improvement over the initial estimate derived from the IMF and cluster age was observed in only one case. This is primarily due to the presence of a hot MS primary, which, in the other five cases, is luminous enough to outshine nearly any WD in the $NUV$ band, resulting in weak constraints on $T_\text{hottest}$. A comparison with the Kroupa prior is presented in \autoref{appendix:ifmr}, \autoref{fig:ifmr_no_excess}.

In two cases---candidates 114 and 121---the cluster turnoff mass exceeds the $8\,\text{M}_\odot$ upper limit for WD progenitors. These systems are of interest, as discussed in \autoref{appendix:excluded_systems}, and their analysis is deferred to a future publication. For candidate 48, the analysis yielded a minimal WD cooling age larger than the cluster age, suggesting that a WD companion should have been detectable in the $NUV$. The most likely explanation for this discrepancy is an underestimation of the cluster age (see \autoref{appendix:excluded_systems}).

\subsection{No UV Data}

The third group in our sample comprises the 16 candidate MS+WD binaries with no UV photometry. For these sources, the lack of UV data prevents us from constraining $T_{\text{eff},2}$ or definitively confirming the nature of the companion. Nevertheless, a WD companion remains more plausible than alternatives such as neutron star or SD companions. Assuming all 16 systems are indeed MS+WD binaries, we assign probability distributions to the corresponding progenitor masses and leave detailed, case-by-case validation for a future work.

\begin{figure*}
    \centering
    \includegraphics[width=0.7\textwidth]{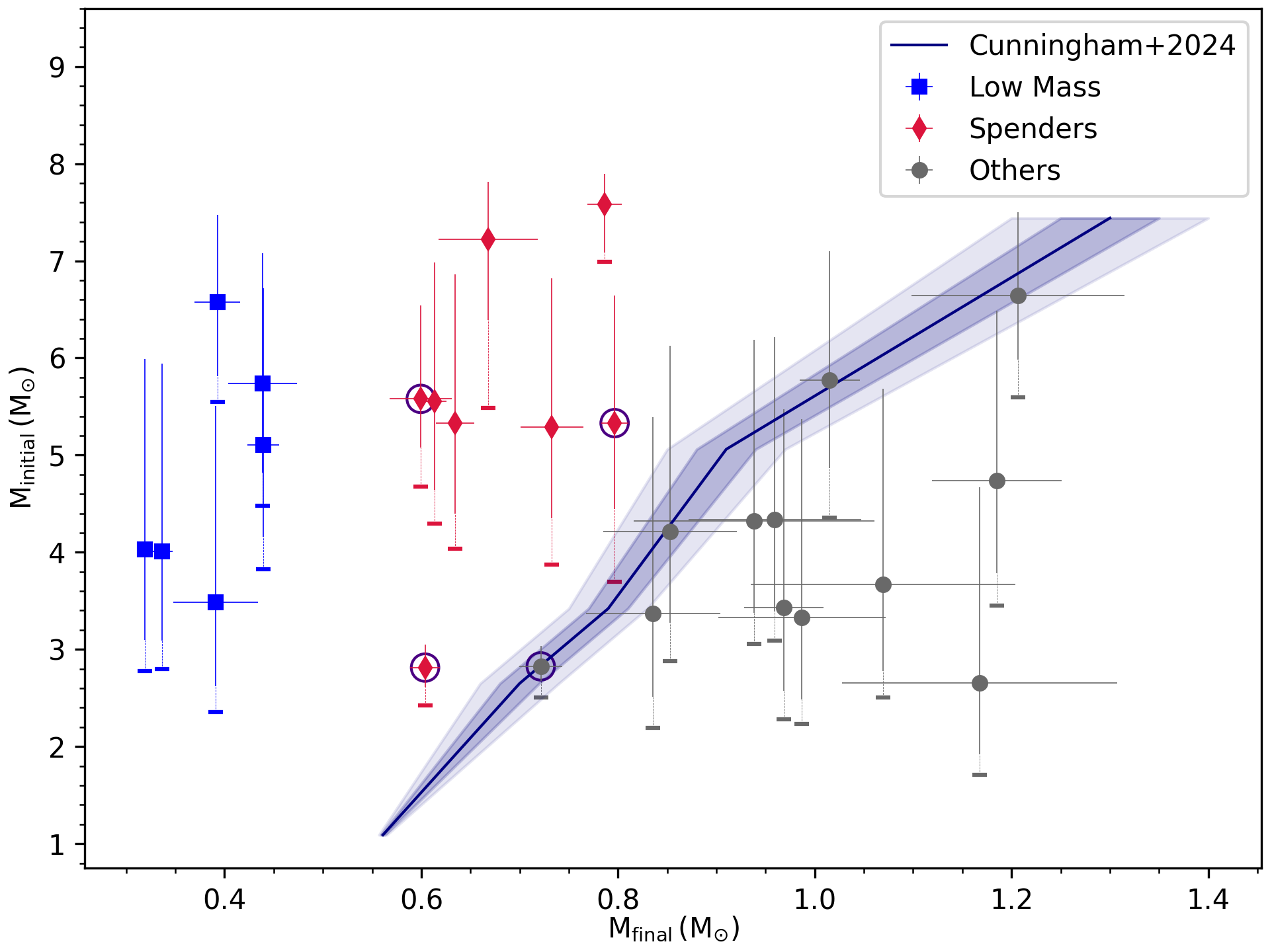}
    \caption{Estimated progenitor mass ($M_\text{initial}$) versus the dynamically measured remnant mass ($M_\text{final}$). Progenitor masses and error bars are defined by the 16\textsuperscript{th}, 50\textsuperscript{th} and 84\textsuperscript{th} percentiles of the $M_\text{initial}$ distribution (see \autoref{sec:ifmr}). The dark blue solid line shows the piecewise linear IFMR from \citet{cunningham2024}, with colored regions indicating the one- and two-sigma confidence intervals. The sample is divided into three groups: ‘low-mass’ WDs, located outside the valid range of the Cunningham IFMR; ‘spender’ WDs, positioned above the line, corresponding to super-nominal mass loss; and ‘others’, systems positioned below the line, apparently consistent with sub-nominal mass loss. The encircled data points highlight the four systems with UV excess (see \autoref{fig:ifmr_uv_excess}). Thick horizontal dashes indicate strict lower limits for the progenitor masses, derived from either the respective cluster MS turnoff mass, or from the posterior of $M_\text{initial}$ for UV excess candidates.}
    \label{fig:ifmr_groups}
\end{figure*}

For each system, we construct a Kroupa prior, similar to \autoref{fig:kroupa_prior}, based on the cluster association, the IMF, and the $8\text{M}_\odot$ upper limit for WD progenitors. The median of each Kroupa distribution is adopted as the best estimate for the progenitor mass, with error bars determined from the 16\textsuperscript{th} and 84\textsuperscript{th} percentiles. The results are shown in \autoref{fig:ifmr_groups}, and discussed in \autoref{sec:discussion}.

\section{Sample Properties}\label{sec:sample_properties}
In this section, we examine the orbital parameters of our final sample, focusing on period, eccentricity, and their relationship with secondary star masses. We first analyze the observed period-mass relation to identify potential signatures of binary interaction. We then evaluate the impact of selection biases in astrometric binaries and their influence on the detected system properties.

\subsection{Period-Mass Relation}

\begin{figure}
    \centering
    \includegraphics[width=\columnwidth]{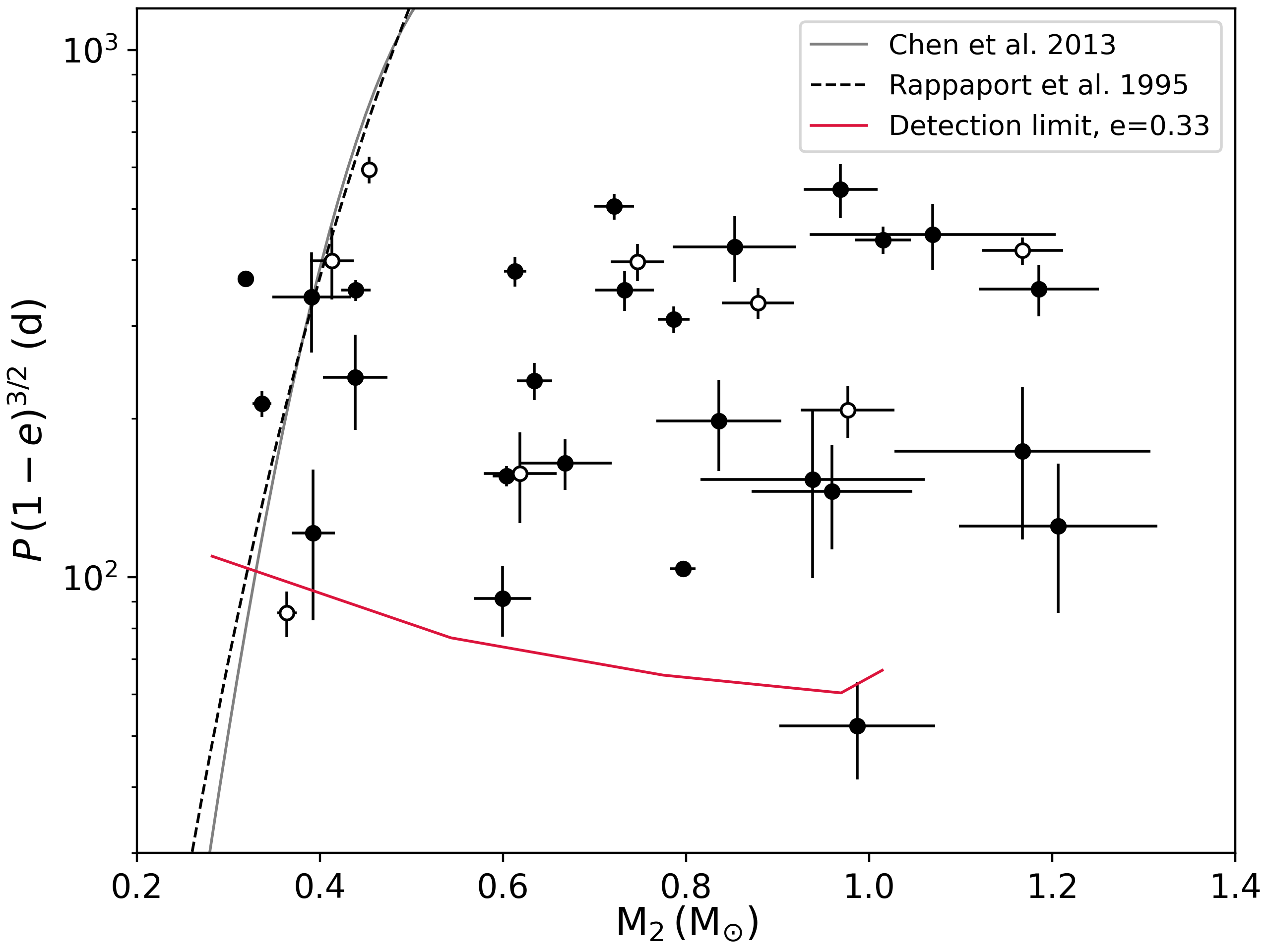}
    \caption{Period-mass relation for 34 systems in our sample. Black markers represent the selected 26 MS+WD binaries, and hollow markers indicate excluded systems (\autoref{appendix:excluded_systems}). The solid black line shows the theoretical
    relation from \citet{chen_pm}, while the dashed gray line shows the relation from \citet{rappaport_pm}. Both relations are calculated for a metallicity Z = 0.02 typical to our sample. The y-axis shows the
    orbital period, $P$, times $(1-e)^{3/2}$, where $e$ is the eccentricity. The entire sample falls roughly between 100-1000 days, due to the period range of NSS astrometric binaries. The red line shows, for different WD masses, the minimal period detectable by the AMRF triage method, using $e=0.33$, the median eccentricity of our sample. Systems having exceptionally high eccentricities are potentially detectable below this line.}
    \label{fig:period_mass}
\end{figure}

\autoref{fig:period_mass} shows the observed period-mass relation for our sample, along with theoretical relations for interactions occurring during the red-giant branch (RGB) phase. The figure presents our sample of candidate post-interaction binaries, including the seven systems excluded due to cluster membership or age uncertainties and our single MS+SD candidate.

Post-binary interaction systems are expected to exhibit a relationship between the orbital period and the WD mass \citep[e.g.,][]{joss_pm}. In the low-mass regime, the theoretical models of \citet{rappaport_pm} and \citet{chen_pm} for post-RGB mass transfer predict that binaries with orbital periods between $100$ and $1000$ days should host WDs with masses around $0.4\,\text{M}_\odot$. As the figure shows, this prediction aligns well with the observed low-mass WDs in our sample. The agreement suggests that these low-mass WDs result from stable mass transfer during their progenitors' RGB phase. 

For the orbital periods considered here, WDs with masses exceeding ${\sim}\,0.5\,\text{M}_\odot$ do not align with the stable mass transfer relations of the RGB phase. This discrepancy suggests that these systems likely underwent alternative evolutionary pathways to account for their higher masses \citep[see, e.g.,][and references therein for potential alternatives, though primarily for systems with shorter orbital periods and lower eccentricities]{Cohen_2024}.

We also note the absence of `extremely low-mass WDs' (ELMs; $M_2 < 0.3\,\text{M}_\odot$) in our sample. This absence is likely due to their shorter orbital periods and lower companion masses, which result in closer orbital separations. Consequently, these systems fall below the detection limits of our method, as shown by the red line in \autoref{fig:period_mass}.

In our final sample (\autoref{fig:ifmr_groups}; \autoref{fig:period_mass}; \autoref{fig:ifmr_parallax}), there appears to be a gap in WD candidates between roughly $0.45$ and $0.55\,\text{M}_\odot$. However, WDs in this mass range were identified in other studies (e.g., \citealt{Shahaf2025}). We are unaware of a specific bias disfavoring their detection. We suggest this gap is a statistical artifact due to the small sample size.

\subsection{Selection Bias}\label{subsec:bias}

Several observational and astrophysical biases influence our sample. It is restricted to Gaia astrometric binaries associated with open clusters, which limits the analysis to relatively young systems with orbital periods of 100–1000 days. Additionally, the AMRF triage method preferentially identifies systems with massive companions. These selection effects impose constraints on the range of orbital separations we are sensitive to. Therefore, given Gaia's finite angular accuracy, the detection bias is expected to scale with the system's distance.

As an order-of-magnitude estimate, consider systems with an observed semi-major axis of $\alpha_0 \gtrsim 1\,\text{mas}$, as typically found in the sample of astrometric MS+WD candidates by \citet{Shahaf_2024}. Assuming orbital separations of ${\sim}\,1\,\text{AU}$, this angular separation limit corresponds to a minimum parallax of $1\,\text{mas}$, or a maximum distance of approximately $1\,\text{kpc}$. Beyond this distance, our sensitivity declines significantly, as non-massive companions struggle to produce a detectable $1\,$mas orbit on the primary. However, massive WDs, which induce larger orbital signals, remain detectable at slightly greater distances. 

A detailed analysis of all selection effects is beyond the scope of this study. However, a simple quantitative estimate of the scale of these effects as a function of distance is feasible. To this end, we define the selection probability ratio, $\mathcal{R}$, as the ratio of the probability of detecting WDs more massive than $0.8\,\text{M}_\odot$ to the probability of detecting WDs below this limiting mass, namely

\begin{equation}\label{eq:odds_ratio}
    \mathcal{R} \triangleq \frac{\sum p(M_2>0.8\,\text{M}_\odot)}{\sum p(M_2\leq0.8\,\text{M}_\odot)}.
\end{equation}
This ratio is quantified using the code developed by \citet{nss_sf}, as outlined in \autoref{appendix:bias}. 

\autoref{fig:bias_vs_distance} illustrates the dependence of $\mathcal{R}$ on cluster distance. While the selection bias remains relatively modest for clusters within $1\,$kpc, it increases substantially beyond this threshold. \autoref{fig:ifmr_parallax} shows the IFMR sample from \autoref{fig:ifmr_groups}, now colored by distance (above or below $1\,$kpc), to emphasize which candidates are most affected by this selection bias.

\begin{figure}
    \centering
    \includegraphics[width=\columnwidth]{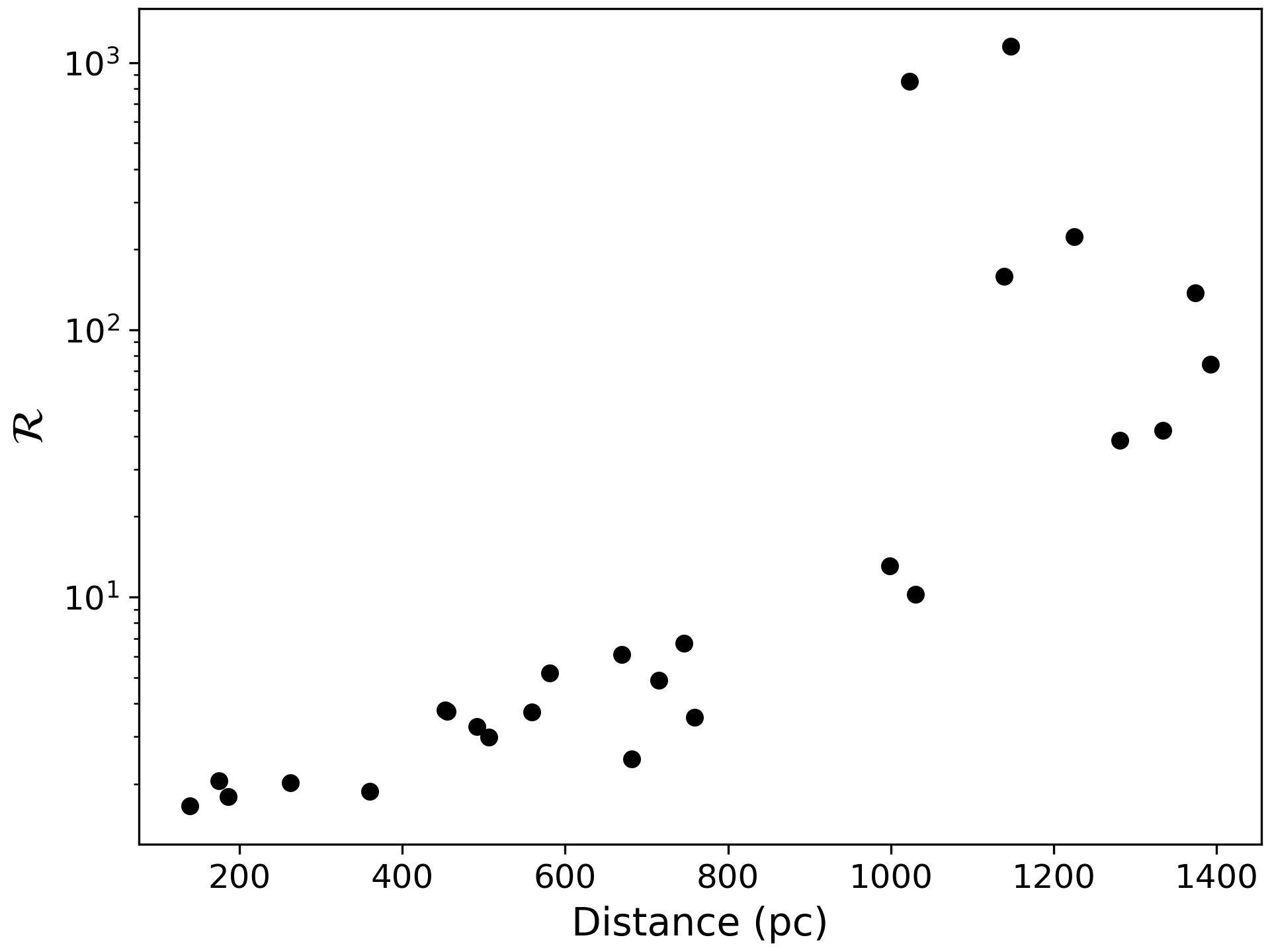}
    \caption{The selection probability ratio $\mathcal{R}$, as defined in \autoref{eq:odds_ratio}, plotted against cluster distance for all 26 systems from \autoref{fig:ifmr_groups}. Selection bias increases significantly beyond $1\,$kpc, with a stronger preference for massive WDs at greater distances.}
    \label{fig:bias_vs_distance}
\end{figure}

\begin{figure}
    \centering
    \includegraphics[width=\columnwidth]{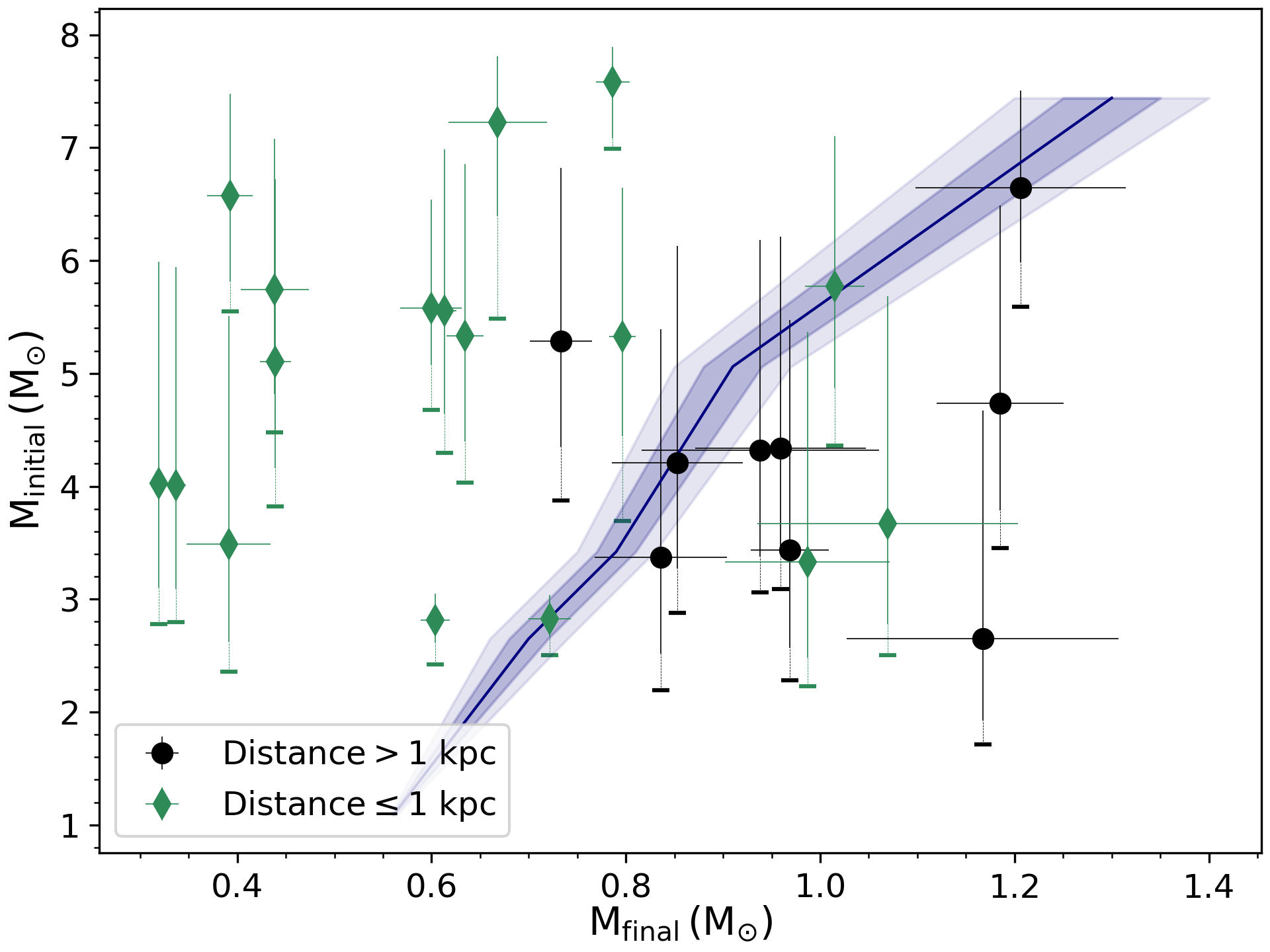}
    \caption{The same IFMR sample as in \autoref{fig:ifmr_groups}, but now with markers and colors indicating distance. Green diamonds represent systems within $1\,$kpc, where selection effects are less prominent. Black circles represent systems beyond $1\,$kpc, where detection is strongly biased toward massive WDs.}
    \label{fig:ifmr_parallax}
\end{figure}

\section{Discussion}\label{sec:discussion}

The analysis presented above yielded estimates for the progenitor masses for 26 WD candidates in binary systems, with orbital separations of  $\sim$1\,AU. \autoref{fig:ifmr_groups} presents our estimate of the progenitor mass, M$_\text{initial}$, versus the dynamical mass estimate of the remnant, M$_\text{final}$. 

The analyzed sample includes 22 candidates where no signature from the WD companion is detectable in the available SED bands (`no UV' or `no UV excess'), and where the progenitor mass is not measured but instead constrained using the IMF prior. Using the prior has been shown to be reliable, as seen in \autoref{fig:ifmr_uv_excess}, provided the system is indeed an MS+WD. Admittedly, some of these systems may not be MS+WD binaries, and case-by-case validation is needed to constrain their true nature. Nevertheless, given the high certainty that the secondary stars in these systems are non-luminous, the MS+WD configuration remains the most probable, with the number of exceptions expected to be of order unity. Our discussion, therefore, focuses on the overall properties of the sample rather than individual systems, ensuring that our conclusions remain robust and independent of a few outliers.

To further support the reliability of the sample, we carefully vetted each system to rule out potential problems with cluster membership or the age estimate. This addressed known issues that affected the excluded systems (\autoref{appendix:excluded_systems}), ensuring they are unlikely to affect the retained systems.

We highlight two distinct populations within our sample, each with a different evolutionary channel: `low-mass' and `spenders'. These groups are defined and characterized below. 

\subsection{Low mass}

The first group consists of 6 candidates with $M_\text{final} \sim 0.4\,\text{M}_\odot$. As shown in \autoref{fig:period_mass}, their observed period-mass relation aligns well with theoretical predictions for systems that underwent complete envelope stripping during the RGB phase \citep{rappaport_pm,chen_pm}. Additional evidence supporting the binary origin of these WDs is presented in \autoref{fig:ifmr_groups}, which illustrates that the classical IFMR breaks down for WD masses below $\sim0.55\,$M$_\odot$. Single-star progenitors of such low-mass WDs could not have evolved off the MS within the Galaxy’s lifetime. Therefore, these WDs are incompatible with single-star evolution. 

Theoretically, these low mass WDs are expected to have a helium core \citep[e.g.,][]{Althaus_2013} or a small carbon-oxygen core surrounded by a thick helium shell \citep[`hybrid' WDs; e.g.,][]{Iben_1985, Zenati_2019, Romero_2022}, as the interaction halts helium burning prematurely during the RGB phase of the progenitor. 

As \autoref{fig:ifmr_parallax} shows, all low-mass WD candidates are found in relatively nearby open clusters. This observation aligns with the conclusion of \autoref{subsec:bias} that there is a significant detection bias against these systems beyond 1\,kpc.

Additionally, two confirmed low-mass WDs (candidates 281 and 283) identified through SED fitting are missing from \autoref{fig:ifmr_groups}, as their progenitor masses could not be derived reliably. See \autoref{appendix:excluded_systems} for further details.

\subsection{Spenders}
The second group comprises eight candidates with $0.6\,\text{M}_\odot < M_\text{final} < 0.8\,\text{M}_\odot$. Our analysis suggests that the progenitors of these WDs are overly massive, compared to what is expected from the \citet{cunningham2024} IFMR. These systems are labeled as ‘spenders', as they appear to have lost more mass than an equivalent single star. This observation is not sensitive to the details of our progenitor mass estimation process, as in most of these cases, the IFMR predictions fall below the lowest possible mass obtained from the turnoff mass of the cluster (see horizontal dashes in \autoref{fig:ifmr_groups}). These candidates align with our expectation of enhanced mass loss in post-interaction systems. Unlike single-star evolution, where mass loss occurs only through stellar winds and envelope ejection, additional mass loss is predicted to be induced by the interaction between the components.

Among MS+WD binaries with periods of $100-1000$~days, the spender group represents a previously unexpected population, as evidenced by its deviation from theoretical predictions in \autoref{fig:period_mass} (also see discussion of similar systems in \citealt{Shahaf_2024}). While the low-mass group is consistent with the \citet{rappaport_pm} and \citet{chen_pm} relations, suggesting interaction during the RGB phase, the spenders may have undergone interaction only during their progenitors' asymptotic giant branch (AGB) phase. We propose this distinct formation mechanism as the cause of the discrepancy between the spenders and theoretical predictions. 

Like the low-mass candidates, the majority of `spenders' reside under 1\,kpc (\autoref{fig:ifmr_parallax}). Their typical mass of $0.6\,\text{M}_\odot < M_\text{final} < 0.8\,\text{M}_\odot$ means it would be significantly more difficult to detect them at greater distances.

Further investigation of these systems could yield valuable insights into binary evolution models, particularly the interplay between parameters such as component masses, orbital separation, metallicity, and other factors that influence mass loss in binary systems. For example, \autoref{fig:ifmr_uv_excess} shows that candidates 175 and 236 share similar properties, including nearly identical present-day component masses ($M_1$ and $M_2$). However, candidate 236, with a shorter orbital period, exhibits a significantly higher initial mass, indicating greater mass loss, consistent with expectations for tighter orbits.

The remaining 12 systems (hereafter referred to as `others'), indicated by gray circles in \autoref{fig:ifmr_groups}, do not fit neatly into either the ‘low-mass’ or ‘spender’ categories. Further observations are needed to determine the formation channel for each of these systems. In contrast, for the ‘low-mass’ and ‘spender’ systems, the inconsistency between their WD or turnoff mass with single-star evolution predictions already provides insight into their formation channel. Generally speaking, the `other' candidates host massive WDs, and are preferentially located at greater distances (\autoref{fig:ifmr_parallax}) where their detection is aided by a considerable selection bias, suggesting that they are rarer. In the following paragraphs, we outline potential formation mechanisms for these massive WD candidates, and propose experiments to differentiate between the various scenarios.

In all but one of the ‘other’ candidates, the progenitor mass is loosely bound, with the IMF prior serving as the only constraint. Thus, for each system, the true $M_\text{initial}$ is permitted to be either above (indicating enhanced mass loss) or below (indicating apparent suppressed mass loss) the isolated WD IFMR in \autoref{fig:ifmr_groups}. To resolve this uncertainty, obtaining UV photometry and measuring the WD cooling age is essential to reduce the initial mass uncertainty. If future observations are consistent with massive progenitors, they would support enhanced mass loss, as expected for post-interaction systems. This would help rule out suppressed mass loss, which we believe to be non-physical.

However, our current constraints allow, and even favor, lower progenitor masses, as seen in \autoref{fig:ifmr_groups}. Consequently, we propose a possible formation channel for massive WDs that involves non-massive progenitors without requiring suppressed mass loss. In single-star evolution, it is estimated that over $20\%$ of isolated high-mass WDs form through double WD mergers, rather than from a single massive progenitor \citep[e.g.,][and references therein]{merger_empirical, merger_pop_syn, mass_deficit}. Accordingly, we suggest that some of the ‘other’ WD candidates may arise from triple system evolution. This configuration involves a tight binary of massive MS stars orbited by a less massive MS star (present primary) at a few AU. The inner binary evolves into either a merged or double WD system (present secondary), with each progenitor undergoing enhanced mass loss during its evolution. The present-day dynamical mass thus reflects the combined masses of the two resulting WDs. This idea is reinforced by the recent discovery of such a system in an old open cluster \citep{Leiner_2025}. This system consists of an MS and a relatively massive WD $\sim 1$\,AU apart, that was identified as a merger product based on its cooling age.

It is thus possible to test this hypothesis through UV photometry and precise measurements of WD cooling ages. A significant mismatch between the cooling age and the cluster age would support the merger/double-WD scenario \citep[e.g.][]{Leiner_2025} over the single-progenitor model. Systems identified in the future as merger products or unresolved double-WDs will require special consideration, as a double progenitor scenario is not accounted for in \autoref{fig:ifmr_groups}. Nevertheless, these descendants of hierarchical triples are of particular interest, as the potential progenitors of type Ia supernovae \citep[and see discussion in][]{Leiner_2025}.

To differentiate between a massive post-merger WD and an unresolved binary of intermediate-mass WDs, we suggest obtaining HST FUV spectra. The equivalent width of absorption lines would be substantially larger in the single massive WD case, reflecting its high surface gravity, while a tight double degenerate binary would produce narrower lines due to the combined spectra of lower-gravity components.

\section{Conclusions}\label{sec:conclusions}
This paper presents a sample of astrometric binaries in open clusters, selected through a systematic process. From an initial dataset of 283 sources, we applied astrometric and SED analyses to identify 33 probable MS+WD binary candidates and estimated progenitor masses for 26 WDs. While individual validation of each candidate will require FUV photometry or spectroscopy, our selection criteria ensure that the sample as a whole provides a reliable basis for studying the IFMR. Additionally, the sample size is expected to grow with Gaia DR4.

The sample reveals two distinct populations based on a comparison with the IFMR of isolated WDs. The first group comprises of low-mass WD candidates, likely helium-core WDs formed through envelope stripping during the progenitors' RGB phase. The second group, consisting of intermediate-mass WDs, shows evidence of an enhanced mass loss compared to classical IFMR predictions, also likely due to binary interaction. Both groups offer valuable cases for studying how factors such as orbital parameters, component masses, and metallicity influence mass loss in binary systems. By analyzing this sample, we can better understand how these factors influence the diversity of WD outcomes in binary systems. 

For the remaining candidates, typically with high-mass WD companions, the evolutionary history of each system remains undetermined. Given the relatively old cluster ages, both massive and non-massive progenitors could plausibly have evolved to form a WD. As a result, current evidence does not allow us to determine whether these progenitors experienced enhanced or suppressed mass loss. Precise measurements of WD cooling temperatures through UV photometry could provide key insights to constrain the true initial masses of these systems.

Furthermore, it is plausible that some of these systems originate from triple-system progenitors, which would necessitate special consideration. In this scenario, the present-day companion is either a double WD or a single massive WD formed through a binary merger. Such a system was recently discovered by \citet{Leiner_2025} in the M67 open cluster. If UV photometry reveals discrepancies between WD cooling ages and single-star evolutionary expectations, it could point to the triple progenitor hypothesis as the favorable formation mechanism. 

These systems are promising targets for follow-up observations due to their high potential impact, as massive WDs in binaries are thought to be type Ia supernova progenitors. Spectroscopy is crucial for validating orbital parameters and confirming the reliability of individual orbits, although studies of similar systems indicate a high success rate for Gaia DR3 orbital solutions \citep{Yamaguchi2024}. Additionally, FUV spectroscopy could differentiate between a single massive WD and a tight binary of intermediate mass WDs by the width of the Ly$\alpha$ line. High-resolution spectroscopic observations could, in some cases, further confirm the binary nature of the faint massive companion or set limits on the orbital separation of the inner binary pair \citep{Nagarajan2024}.

The recently discovered population of `Blue Lurkers' \citep{Leiner_2019}, lower-luminosity counterparts to the blue straggler stars, identified by significantly shorter rotation periods compared to normal MS stars, resembles the MS+WD sample presented here. Similar to our sample, blue lurkers also host WD companions at orbital separations of $\sim$1\,AU, and show evidence for a past mass transfer event \citep[e.g.][]{Nine_2023, Jadhav_2024}. It is likely that both samples are part of the same population. Measurements of the rotation periods of some of the MS stars in our sample (through photometric variations or spectral line broadening) might reveal indications for a spin-up compared to single MS stars of the same age.

Of the eight systems without assigned progenitor masses, one is likely an MS+SD binary, while two others show a significant tension between the cluster age and WD cooling age. We also identified four systems where the cluster turnoff mass exceeds 8,M$_\odot$. If our future studies confirm the reliability of the age estimates, it would indicate that the nature of the compact companion is intriguing. These cases are excellent candidates for future studies on WD evolution in binaries.

\section{Acknowledgments}

We thank Boaz Katz, Doron Kushnir, Tsevi Mazeh, Eli Waxman, and the anonymous referee for valuable comments.
SBA acknowledges support from the Israel Ministry of Science grant IMOS 714193-02.
SBA's research is supported by the Peter and Patricia Gruber Award; the Azrieli Foundation; the Andr\'e Deloro Institute for Advanced Research in Space and Optics; the Willner Family Leadership Institute for the Weizmann Institute of Science; and the Israel Science Foundation grant ISF 714022-02.
SBA is the incumbent of the Aryeh and Ido Dissentshik Career Development Chair.
SS is supported by a Benoziyo prize postdoctoral fellowship.

This work has made use of data from the European Space Agency (ESA) mission
Gaia (\url{https://www.cosmos.esa.int/gaia}), processed by the Gaia
Data Processing and Analysis Consortium (DPAC,
\url{https://www.cosmos.esa.int/web/gaia/dpac/consortium}).

This publication makes use of data products from the Wide-field Infrared Survey Explorer, which is a joint project of the University of California, Los Angeles, and the Jet Propulsion Laboratory/California Institute of Technology, funded by the National Aeronautics and Space Administration.

This publication makes use of data products from the Two Micron All Sky Survey, which is a joint project of the University of Massachusetts and the Infrared Processing and Analysis Center/California Institute of Technology, funded by the National Aeronautics and Space Administration and the National Science Foundation.

All the GALEX data used in this paper can be found in MAST: \dataset[10.17909/8h2g-ay89 ]{http://dx.doi.org/10.17909/8h2g-ay89 }. 

Funding for the DPAC
has been provided by national institutions, in particular the institutions
participating in the Gaia Multilateral Agreement.
This research has made use of the Spanish Virtual Observatory (\url{https://svo.cab.inta-csic.es}) project funded by MCIN/AEI/10.13039/501100011033/ through grant PID2020-112949GB-I00.
\vspace{5mm}
\facilities{Gaia, GALEX, 2MASS, WISE}

\software{astropy \citep{Astropy_2013, Astropy_2018}, astroquery \citep{astroquery}, emcee \citep{emcee}, matplotlib \citep{Hunter_2007}, numpy \citep{Numpy_2006, Numpy_2011}, scipy \citep{Virtanen_2020}, topcat \citep{Taylor_2005}, dustapprox \citep{Fouesneau_dustapprox_2022}}

\appendix
\section{Sample selection}\label{appendix:sample}
We use two independent open-cluster catalogs with Gaia membership lists, cross-checking one against the other and expanding the sample size with non-exclusive clusters.
A schematic description of the sample selection (along with the data analysis, see \autoref{sec:analysis}) is presented in \autoref{fig:full_flowchart}.

\subsection{Hunt \& Reffert's Catalog}
The \citet{Hunt_2023} catalog, based on Gaia DR3, focuses on constructing robust membership lists for as many clusters as possible. This study identifies 7167 open clusters, including 2387 newly reported candidate clusters. For our analysis, we use a subset of 4105 highly reliable clusters recommended by the authors. Among these, approximately 600 candidate clusters may represent gravitationally unbound moving groups. However, since these groups also constitute single-age populations, we have included them in our sample nonetheless. The robust subsample from \citeauthor{Hunt_2023} contains 658,507 Gaia-DR3 members. From this subset, we apply the following selection criteria: 
\begin{itemize}
    \item $\text{parallax}/\text{parallax error} > 10$ and
    \item and $\text{Astrometric fidelity} > 0.5$.
\end{itemize}
For details regarding the definition of astrometric fidelity, see \citet{Rybizki_fidelity}.

\subsection{Dias et al. Catalog}
The Gaia-DR2-based catalog of \citet{Dias_2021} is aimed to provide a homogeneous and accurate sample of cluster parameters. The catalog borrows membership lists from previous works, in some cases recalculating membership probabilities. The CMDs of 1743 clusters were fit with PARSEC isochrones, looking to determine the best solutions for the cluster age, metallicity, extinction and distance. We apply the same parallax-over-error and astrometric fidelity cuts once more, this time to the \citeauthor{Dias_2021} membership lists.

\subsection{Gaia NSS Catalog}
The Gaia NSS catalog \citep{gaia_nss}, released in 2023 as part of the third data release \citep{gaiadr3}, presents solutions for 800,000 astrometric, spectroscopic, and eclipsing binary systems identified by the Gaia mission. Among these, approximately 150,000 are astrometric binaries, where the photo-center of the source exhibits a discernible wobble around the anticipated motion for a single star. This additional motion may indicate the presence of a companion, either significantly fainter than the primary or too close to be resolved.

Our analysis focuses exclusively on systems categorized with NSS solution types \texttt{Orbital} or \texttt{AstroSpectroSB1}---retaining astrometric binaries but discarding purely spectroscopic/eclipsing systems. For these systems, the catalog provides the essential orbital parameters required for our investigations (Thiele-Innes A, B, F, G parameters and the orbital period, as described in \citealt{nss_astrometric}), and they are deemed more reliable than solutions labeled \texttt{OrbitalAlternative}. 

To avoid spurious solutions, we take the cuts recommended by \citet{nss_astrometric} on the orbital solution:
\begin{itemize}
    \item Eccentricity error, $\delta e <0.079\log(P/\text{day}) -0.244$
    \item Parallax over error, $\varpi/\delta \varpi > 20,000 (P/\text{day})^{-1}$
    \item Significance, $s>\max\left(5,~158/\sqrt{P/\text{day}}\right)$
\end{itemize}
where $P$ is the orbital period.

\subsection{Cross-Matching}\label{subsec:xmatch}
We are only interested in NSS sources that are highly confident cluster members. As such, we only consider candidates with a membership probability greater than $90\%$. The 4105 high-confidence clusters of \citet{Hunt_2023} contain 562 unique NSS sources with $>90\%$ membership probability, associated with 428 different open clusters. Furthermore, only 212 sources have the appropriate NSS solution type (\texttt{Orbital}/\texttt{AstroSpectroSB1})---and are associated with 186 open clusters. In 1743 \citet{Dias_2021} clusters, we find 408 NSS sources in 206 different clusters. Of these, 103 sources in 64 clusters have the required types of NSS solution. There are 32 sources that overlap between the two samples, resulting in a combined total of 283 unique sources spread across 218 distinct open clusters. For each source, we collected Gaia-DR3 astrometry, photometry, and NSS orbital parameters (period, Thiele Innes parameters, etc.). We took the parallax and proper motion values reported in the NSS catalog, since in the Gaia-DR3 source table these parameters assume a single body astrometric solution. 

\begin{figure}
    \centering
    \includegraphics[width=\columnwidth]{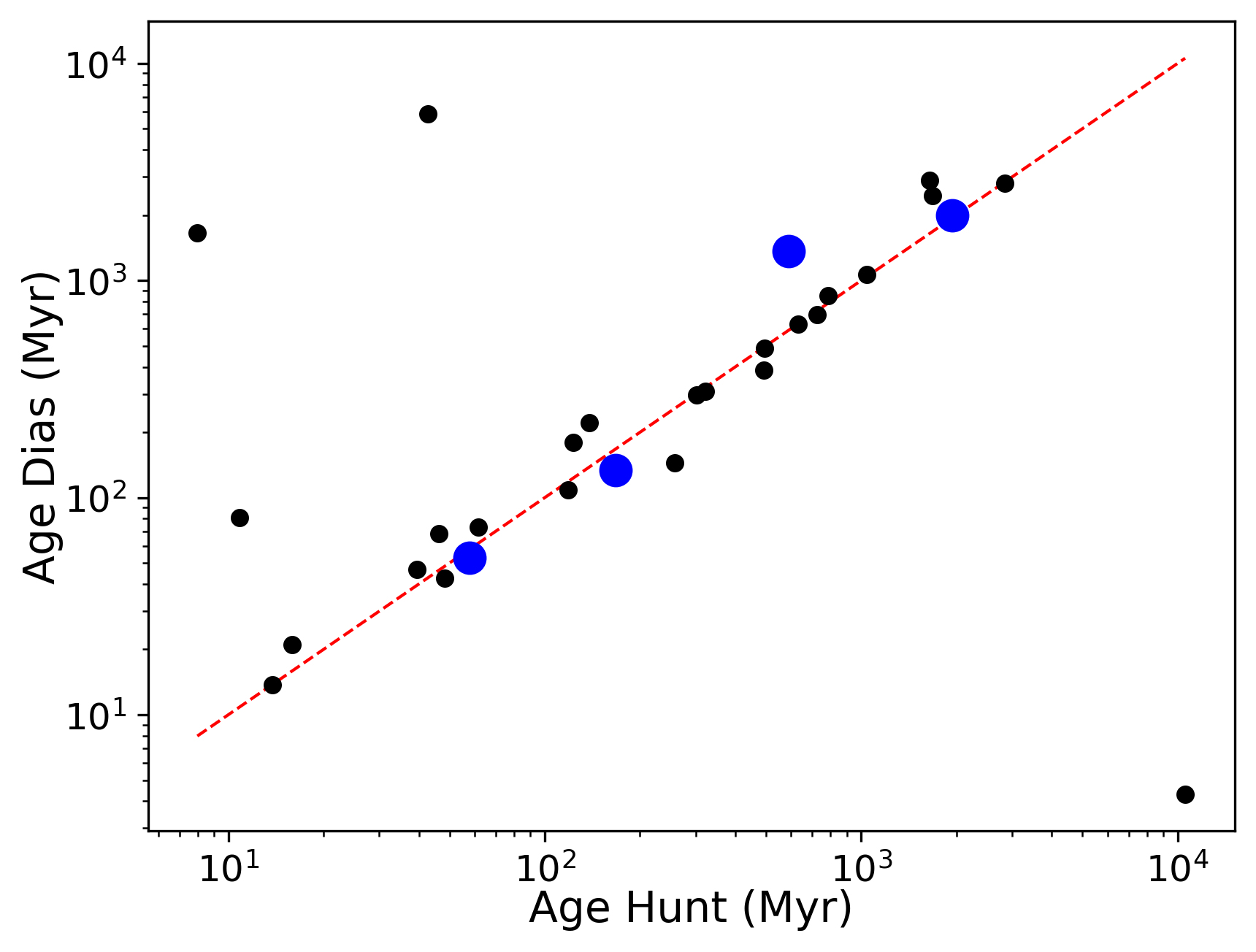}
    \caption{Ages derived using \citet{Hunt_2023} membership lists vs. \citet{Dias_2021} membership lists, for the 30 duplicate clusters appearing in the NSS cross-match. The large blue markers highlight four clusters that enter the final sample of MS+WD candidates.}
    \label{fig:cluster_comparison}
\end{figure}

In \autoref{fig:cluster_comparison}, we present the ages derived using the membership lists from \citet{Hunt_2023} and \citet{Dias_2021} for duplicate clusters appearing in the NSS cross-match.

\section{Extinction} \label{appendix:extinction}
In this work, we report the extinction $A_V$ and reddening $E(B-V)$ in the Johnson system. However, we often use other passbands (like the Gaia bands, GALEX bands, etc.) and need a consistent way to infer from $A_V$ the extinction in these bands, $A_X$. This is done through the extinction coefficients:
\begin{equation}\label{eq:extinction_coef}
    k_X = \frac{A_X}{A_V}
\end{equation}
$A_X$ arises from a complex interplay between the SED $F_\lambda$, the filter transmission curve $T_\lambda$, and the extinction curve $\tau_\lambda$ (which depends on both the normalization $A_0$ and the extinction law). Throughout this work, we use the \textsc{Python} package \textsc{dustapprox} of \citet{Fouesneau_dustapprox_2022}, recommended on the Gaia website. This tool provides out-of-the-box polynomial fits for $k_X$ based on the \citet{F99} extinction law with $R_0=3.1$, and the \citet{Kurucz_Castelli} atmospheric models integrated over various passbands: GALEX $NUV$/$FUV$, Gaia DR3, Johnson $UBVRI$, SDSS $ugriz$, 2MASS $JHKs$, and WISE $W1/W2/W3/W4$.

\textsc{Dustapprox} uses the following parameterization for the extinction coefficients:
\begin{equation}
    \begin{split}
        &\tilde{k}_X(x,y) = \, c_0 + c_x x + c_y y + c_{xx}x^2 + c_{xy} xy \\
        & + c_{yy}y^2 + c_{xxy}x^2 y + c_{xyy}xy^2 + c_{xxx}x^3 + c_{yyy}y^3
    \end{split}
\end{equation}
where $x = A_0$ and $y = T_{\text{eff}}/5040\,\text{K}$. Unlike our \autoref{eq:extinction_coef}, \textsc{dustapprox} normalizes the extinction coefficients with respect to the monochromatic $A_0$:
\begin{equation}
    \tilde{k}_X = \frac{A_X}{A_0}
\end{equation}
To simplify, we calculated the average $\tilde{k}_V$ over the grid:
\begin{equation}
    \langle\tilde{k}_V\rangle \equiv \frac{A_V}{A_0} = 1.02 \pm 0.02
\end{equation}
Then, for any other band, we could easily use the polynomials provided by \textsc{dustapprox} to get the $k_X$ coefficients normalized to $A_V$ as in \autoref{eq:extinction_coef}:
\begin{equation}
    k_X(A_V,y) = \tilde{k}_X(A_0,y) \frac{A_0}{A_V} = \tilde{k}_X\left(\frac{A_V}{1.02},y\right)\times 1.02
\end{equation}
In our work, we choose to redden models, since the model $y=T_{\text{eff}}/5040K$ is known exactly. This approach is technically more accurate than dereddening observations, though the difference is likely negligible. When dereddening observations, the extinction coefficients depend on the measured $T_{\text{eff}}$, and vice versa—one has to know $T_{\text{eff}}$ in order to calculate the extinction coefficients.

\section{Isochrone Fitting} \label{appendix:isochrones}
Throughout this work, we repeatedly use the age, extinction and metallicity derived in \autoref{subsec:cluster_analysis} by fitting the cluster CMDs to isochrone models. However, this must be done with care, as various factors hamper quality isochrone fitting: degeneracies, photometric errors, blue straggler stars, outliers, rarity of massive stars, and more. In this section we present in detail our fitting procedure which slightly modifies the code of \citet{Dias_fitting_code} (also used in \citealt{Dias_2021}) to obtain age, metallicity, and extinction estimates for all 218 open clusters in our sample. The fitting procedure is as follows: 
\begin{itemize}
    \item Assign weights to photometric data: $W_i = P_i\cdot G_{\text{BP},\text{min}}/G_{\text{BP},i}$. That is, weights are higher for bluer and brighter stars in the cluster; and are also proportional to the membership probability, $P_i$. 
    \item Draw 500 isochrone parameters from the prior.
    \item Calculate the weighted likelihood for each set of parameters given the data.
    \item Keep the $10\%$ sets of parameters with the highest likelihood, and use them to derive a new parameter distribution.
    \item Use the new distribution to draw a new sample of 500 isochrone parameters.
    \item Repeat until convergence, or until a maximal number of iterations is reached, reporting the best fit for $\log(t_\text{cluster}),~D,~[\text{Fe}/\text{H}],~A_V$ based on this loop, where $t_\text{cluster}$ is the cluster age, and $D$ is the cluster distance.
\end{itemize}
This process is iterated for 10 total runs, each time taking only a random subset of the observed members as the data (bootstrapping). The final results for each parameter are the mean of these 10 runs, and the reported uncertainty is either the standard deviation or the age/metallicity grid spacing, whichever is larger. There are minor differences between our work and the work of \citet{Dias_fitting_code}, as detailed below:
\begin{itemize}
    \item The \citet{Dias_fitting_code} code was written for Gaia-DR2 data, correcting for systematics in both photometry and parallax. We removed those corrections, which are not relevant for Gaia DR3.
    \item Before fitting, the code discards some members due to low photometric signal-to-noise ratio or the $G_{\text{BP}} - G_{\text{RP}}$ excess factor deviating from its normal range. Members that were disqualified based on DR2 data may participate in the new fit thanks to the improved photometric data in DR3.
    \item We use a $0.02$\,dex grid in metallicity, rather than a  $0.002$\,dex one as the grid spacing rarely limits the uncertainty in the original work (only $2\%$ of clusters have metallicity error less than $0.02\,$ dex).
    \item The code uses extinction polynomials calculated with Gaia-DR2 passbands. We implemented extinction polynomials calculated with the updated Gaia-DR3 passbands, as detailed in \autoref{appendix:extinction}.
\end{itemize}
The prior on the distance $D$ is calculated using Gaia parallaxes, providing a tight constraint on the best-fitting distance. The moderately restrictive $[\text{Fe}/\text{H}]$ prior is a Gaussian distribution, determined from the Galactic gradient, and a standard deviation of $0.1$\,dex. The slightly restrictive prior on $A_V$, also a Gaussian distribution, is estimated from the individual member extinctions obtained from the \citet{stilism_2019} extinction maps. Finally, we take a flat prior on $\log(t_\text{cluster})$ between 6.6 and 10.1. Here and throughout this work, we use the PARSEC 1.2S models, with $6.6<\log(t_\text{cluster}) <10.1$ in steps of $0.05$\,dex and $-0.9 \leq [\text{Fe}/\text{H}] \leq 0.7$ in steps of $0.02$\,dex.

\section{Photometric Mass Estimate}\label{appendix:photometric_mass}

\begin{figure}
\includegraphics[width=\columnwidth]{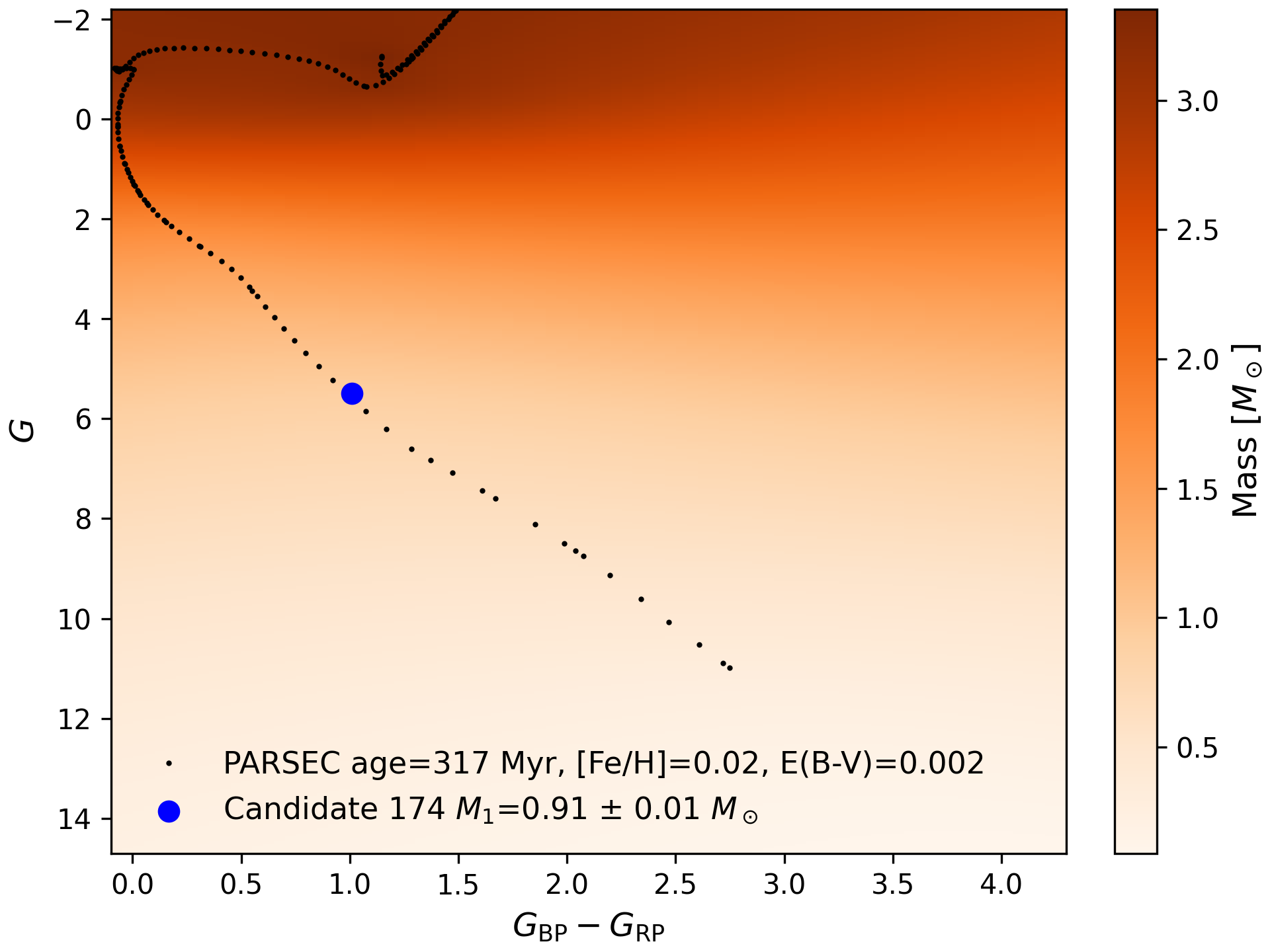}
\caption{Interpolation of the observed color and magnitude along a reddened PARSEC isochrone allows us to estimate the mass of the primary MS star.}
\label{fig:fig1_interpolation}
\end{figure}

The mass is obtained as follows: For each candidate, we generate a PARSEC isochrone with the same age and metallicity as the open cluster it is associated with. The model gives a unique color and magnitude for a given mass. Next, we use the cluster extinction, $A_V$, to estimate $A_G$, $E(G_{\text{BP}}-G_{\text{RP}})$ and apply the extinction to the model isochrone. Lastly, we perform a linear two-dimensional interpolation of the isochrone to get the appropriate $M_1$ for the observed location of the candidate on the Gaia CMD (see \autoref{fig:fig1_interpolation}).

This process is repeated $100$ times in a Monte-Carlo simulation, each time randomly drawing $G$ and $G_{\text{BP}}-G_{\text{RP}}$ from Gaussian priors, taking into account the observational errors. The uncertainties on the cluster parameters are treated the same way: drawing 100 realizations of age, $[\text{Fe}/\text{H}]$ and $A_V$ and generating for every realization an appropriate reddened isochrone. Each iteration produces a different primary mass, $M_1$, with the standard deviation quantifying our uncertainty, $\delta M_1$.

\section{AMRF Triage} \label{appendix:amrf}
In \autoref{subsec:amrf_analysis}, we discuss the upper limit of the astrometric mass ratio function $\mathcal{A}_{\text{MS}}$ for MS+MS binaries. Lacking accurate estimates of the age, extinction, and metallicity,  \citet{Shahaf_2023} used conservative upper limits on $\mathcal{A}_{\text{MS}}$. As a result, some MS+WD systems could be wrongfully classified as  MS+MS binaries. However, since the sample considered in this work has reliable measurements of the age, $A_V$ and $[\text{Fe}/\text{H}]$, we can derive  $\mathcal{A}_{\text{MS}}$ on a case by case basis. For a given $M_1$, these parameters affect the exact value of the MS+MS upper limit, which is determined by the following procedure:
\begin{itemize}
    \item Generate a model isochrone with the appropriate age and metallicity.
    \item From the isochrone, get the effective temperature, $T_{\text{eff},1}$, and radius, $R_1$, of a star with mass $M_1$.
    \item Calculate the flux of a blackbody of radius $R_1$ at a temperature $T_{\text{eff},1}$.
    \item Integrate over the Gaia $G$ band, and apply extinction.
    \item For different $M_2$ values between $0.1-10\,M_\odot$ similarly find $T_{\text{eff},2}$ and $R_2$, integrate the blackbody flux over the $G$ band and apply extinction to get $G_2$.
    \item For each $q=M_2/M_1$ calculate the flux ratio, $\mathcal{S} = G_2/G_1$, and use it to estimate $\mathcal{A}$ (\autoref{eq:amrf_variables}).
    \item Out of all $q$'s, record the maximal $\mathcal{A}$ as the MS upper limit $\mathcal{A}_{\text{MS}} =\mathcal{A}_{\text{max} }(q) $.
\end{itemize}
We run a Monte-Carlo simulation with $10,000$ iterations, each time drawing the orbital parameters $\varpi\,,P\,,\alpha_0$ and $M_1$ from Gaussian priors and calculating $\mathcal{A}$ from \autoref{eq:amrf_obs}. The fraction of instances where $\mathcal{A} <\mathcal{A}_{\text{MS}} $ is denoted `\texttt{class-I probability}'. 
WD companions to MS primaries with $M_1\sim1\,M_\odot$ generally produce $\mathcal{A}$ values that are larger than the MS upper limit $\mathcal{A}_{\text{MS}}$, but not exclusively. We select systems with a `\texttt{class-I probability}' of less than $10\%$. By discarding the majority of the probable MS+MS binaries, which are the primary contaminants of the sample, we also likely exclude some valuable MS+WD systems. As a result, we obtain a smaller, but much cleaner pool of candidates.

\begin{figure}
    \centering
    \includegraphics[width=\columnwidth]{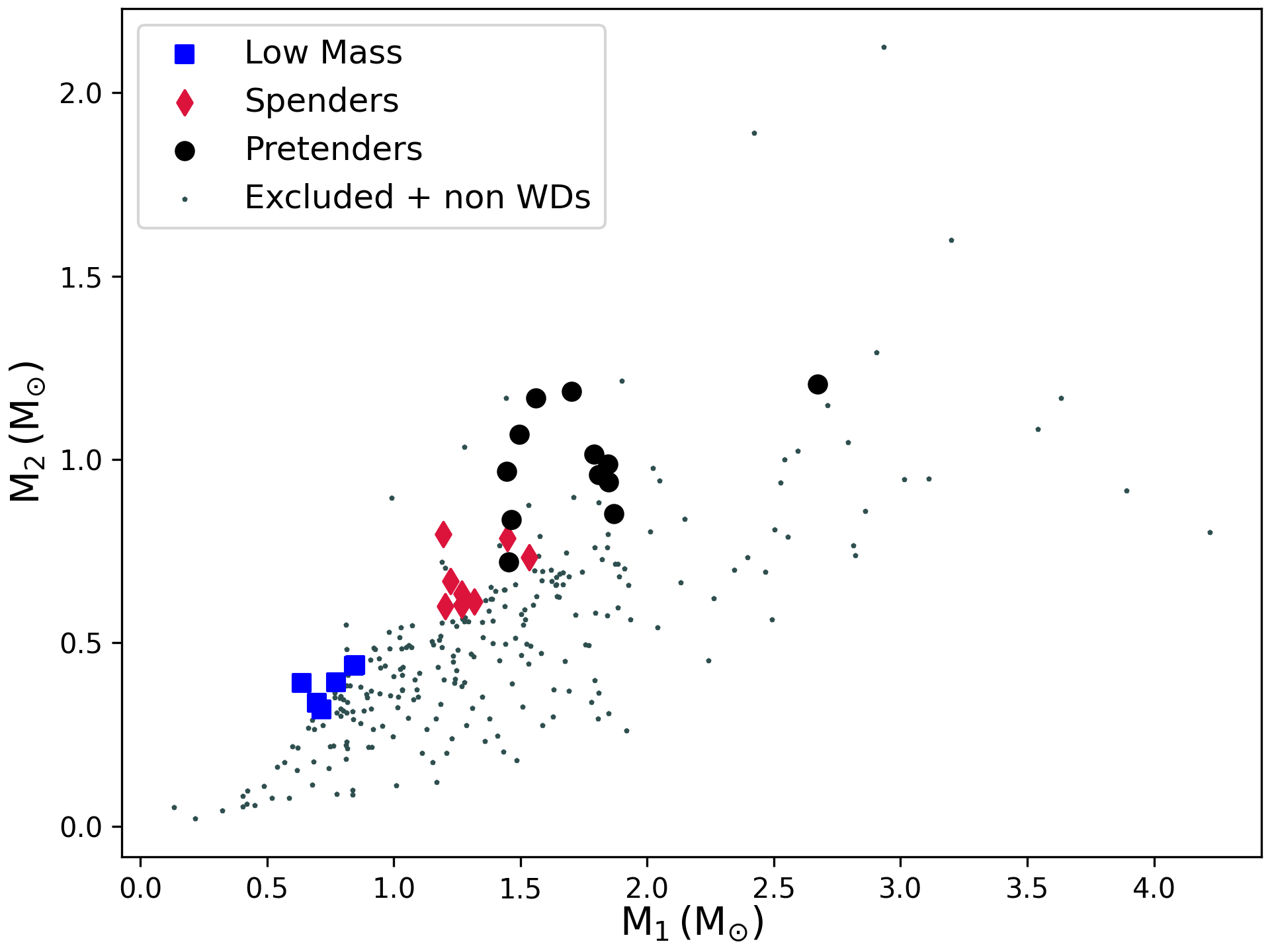}
    \caption{Photometric primary mass vs. companion dynamical mass, both calculated assuming the primary dominates the $G$-band. In blue squares, red diamonds and black circles we show $M_1$ and $M_2$ for the final sample of $26$ MS+WD candidates, for which we present progenitor mass estimates (\autoref{fig:ifmr_groups}). In small gray pentagons we show the remaining sources, including 8 excluded MS+WD candidates and 250 systems identified as likely MS+MS/hierarchical triples.}
    \label{fig:m1_v_m2}
\end{figure}

As mentioned in \autoref{subsec:amrf_analysis}, at this point in the analysis we calculate for each source the companion mass $M_2$, under the assumption that it is much fainter than the primary. In \autoref{fig:m1_v_m2} we plot the photometric primary mass $M_1$ versus the companion mass $M_2$, for the entire sample of $283$ astrometric binaries in open clusters. It is important to note that our derived $M_1$ and $M_2$ accurately reflect the true values only if the primary is a MS star, and the companion is non-luminous in the $G$-band, as is the case for our final sample of MS+WD candidates.

\section{SED Fitting}\label{appendix:sed}
\subsection{Photometric Data}
We construct the SED using the following photometric data: 
\begin{itemize}
    \item In the UV: GALEX $FUV$ and $NUV$ bands.\footnote{We found 18 sources on the second and third General Releases (GR2-GR3) catalog available in the Mikulski Archive for Space Telescopes \citep[MAST;][]{galex-mast} Six of those were also featured on the GR6-GR7 catalog provided by \citet{galex7}; in these cases we opted to use the GR6-GR7 data, though differences were mostly negligible.}
    \item In the optical: Gaia $G$, $G_{\text{BP}}$, $G_{\text{RP}}$ and Gaia synthetic photometry in Johnson-Kron-Cousins $UBVRI$ bands \citep{synthetic_photometry} taking note of synthetic photometry flags (1 for good fluxes, 0 for bad fluxes).
    \item In the IR: 2MASS $J$, $H$, $Ks$ and WISE $W1$, $W2$, $W3$ while respecting the reported quality flags, only keeping measurements flagged as `A' ($\text{SNR}>10$) for 2MASS and `A'/`B' ($\text{SNR}>3$) for WISE.
\end{itemize}
We used the zero points reported in the Spanish Virtual Observatory \citep{SVO_filter_service} to convert magnitudes to flux in physical units (erg\,s$^{-1}$\,cm$^{-2}\,\AA^{-1}$; AB for GALEX, Vega for the visible and IR bands).
As mentioned in \autoref{subsec:sed_analysis}, we take a minimum of $10\%$ uncertainty in all photometric measurements. The original uncertainties are retained for some IR and UV data, where experimental errors are relatively larger.

\subsection{SED Fit to Single Star}\label{subsec:optical_ir_appendix}
For reasons discussed in \autoref{subsec:sed_analysis}, we first fit the SED with a single MS star model. Moreover, we initially use only the optical bands as depicted in \autoref{fig:kurucz_fit_no_excess}. After ruling out IR excess, we include also the IR bands as seen in \autoref{fig:kurucz_fit_with_excess}.

\begin{figure}
    \centering
    \resizebox{\columnwidth}{!}{%
        \begin{tikzpicture}[node distance=1.5cm]
            \node (optical_fit) [process] {Optical SED fit};
            \node (av) [input, above right of=optical_fit, xshift=2cm, yshift=1cm] {$A_V$};    
            \node (m1) [input, above of=optical_fit, yshift=1cm] {$M_1$};
            \node (feh) [input, above left of=optical_fit, xshift=-2cm, yshift=1cm] {[$\mathrm{Fe/H}$]};
            \node (ir_excess) [decision, below of=optical_fit, yshift=-0.5 cm] {IR excess};
            \node (hierarchical_triple) [end, left of=ir_excess,xshift=-2.5 cm] {Hierarchical triple};
            \node (optical_ir_fit) [process, below of=ir_excess,yshift=-0.5cm] {Optical+IR fit};
            \node (uv_excess) [decision, below of=optical_ir_fit, xshift=0cm, yshift=-0.5cm] {UV excess};
            \node (prior1) [output, left of= uv_excess,xshift = -2cm]{$T_{\mathrm{eff},1}, R_1$ prior};
            \node (prior2) [output, below of=uv_excess,yshift=-0.5cm] {$T_{\mathrm{eff},2}, R_2$ prior};
            \node (m2) [input, right of=uv_excess, xshift=2cm] {$M_2$};
            \node (mr_relation) [input, right of=prior2, xshift=2cm] {M--R relation};
    
            \node (full_fit) [process, below of=prior2] {Full SED binary fit};
    
            \node (twd) [end, below of=full_fit] {$T_{\mathrm{WD}}$};
            
            \draw [arrow] (av) -- (optical_fit);
            \draw [arrow] (m1) -- (optical_fit);
            \draw [arrow] (feh) -- (optical_fit);
            \draw [arrow] (m2.south) -- (mr_relation);
            \draw [arrow] (mr_relation.west) -- (prior2);
            \draw [arrow] (optical_fit) -- (ir_excess);
            \draw [arrow] (ir_excess) -- (optical_ir_fit) node[mylabel] {No};
            \draw [arrow] (ir_excess) -- (hierarchical_triple) node[mylabel] {Yes};
            \draw [arrow] (optical_ir_fit) -- (uv_excess);
            \draw [arrow] (optical_ir_fit) -- (prior1);
            \draw [arrow] (prior1) -- (full_fit.west); 
            \draw [arrow] (uv_excess) -- (prior2) node[mylabel] {Yes};
            \draw [arrow] (prior2) -- (full_fit);
            \draw [arrow] (full_fit) -- (twd);
            
        \end{tikzpicture}}
    \caption{SED analysis workflow visualised. The legend is similar to \autoref{fig:full_flowchart}, but here the gray boxes are used for outputs instead of sample size.}
    \label{fig:sed_flow}
\end{figure}
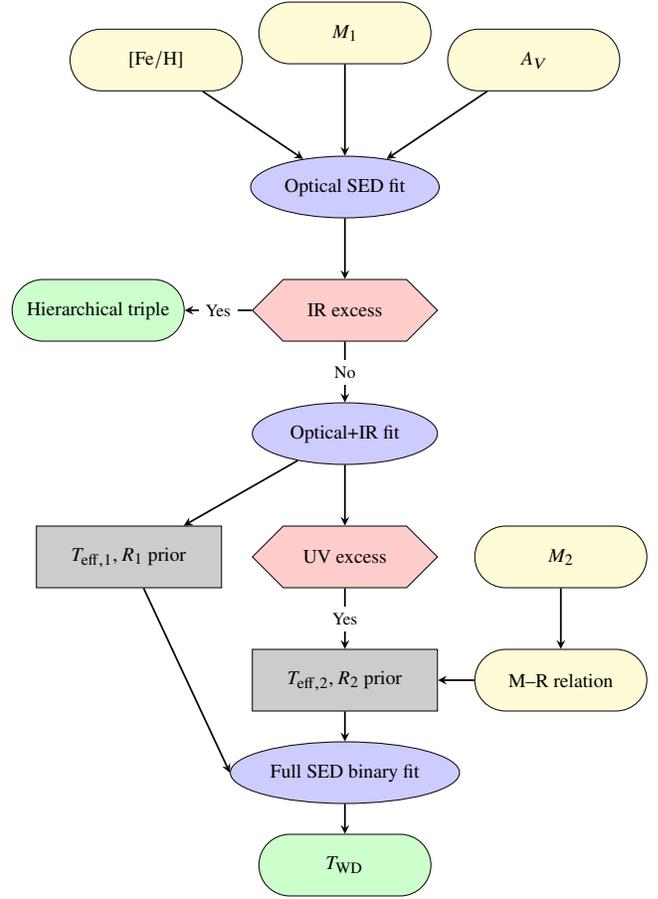

In principle, the single-object fit involves six parameters. Three parameters define the model spectrum $H_\lambda$: the effective temperature ($T_{\text{eff}}$), surface gravity ($\log g$), and metallicity ($[\text{Fe}/\text{H}]$), with each combination producing a distinct model. Two additional parameters are introduced by the scaling factor $(R/D)^2$, where $R$ is the stellar radius and $D$ is its distance. The final parameter is the extinction, $A_V$, which is used to correct the resulting model SED\footnote{In our work, $\lambda$ in the equation below refers to some passband and not a monochromatic wavelength---i.e., $f_\lambda$ is the integrated model flux and $k_\lambda$ is the extinction coefficient in the $\lambda$ band.}: \begin{equation*}
    f_\lambda = 10^{-0.4\cdot k_\lambda A_V}\left(\frac{R}{D}\right)^2 H_\lambda\left(T,~\log g,~[\text{Fe}/\text{H}]\right) 
\end{equation*}\par To reduce the degrees of freedom, we apply prior constraints: the distance $D$ is provided by Gaia, while $[\text{Fe}/\text{H}]$ and $A_V$ are constrained using data from the host cluster (\autoref{subsec:cluster_analysis}), and $M_1$ is estimated from the location on the CMD (\autoref{subsec:mass_analysis}). The remaining parameters, $T_{\text{eff},1}$ and $R_1$ (equivalent to $\log g$ for a fixed mass) are obtained using $\chi^2$ minimization.

\subsection{UV Excess}\label{subsec:uv_excess_appendix}

For the 18 candidates with GALEX photometry, we subtract from the observed flux the $NUV$/$FUV$ contribution of the best-fitting MS model. In nine cases, there is residual flux, which we assume to originate from the companion. Five different parameters ($T_{\text{eff}},~\log g,~R,~D,~A_V$) are required to fit the observed WD flux with WD atmospheric models. Prior knowledge allows us to fix all but one degree of freedom, $T_{\text{eff}}$: We take $D$ from Gaia, and redden the model flux using the cluster's $A_V$. We fix $\log g$ and $R$ by exploiting prior knowledge of $M_2$ from the AMRF (\autoref{subsec:amrf_analysis}), in tandem with a WD mass-radius relation (see \autoref{tab:massradius} for details).

\begin{figure}
    \includegraphics[width=\columnwidth]{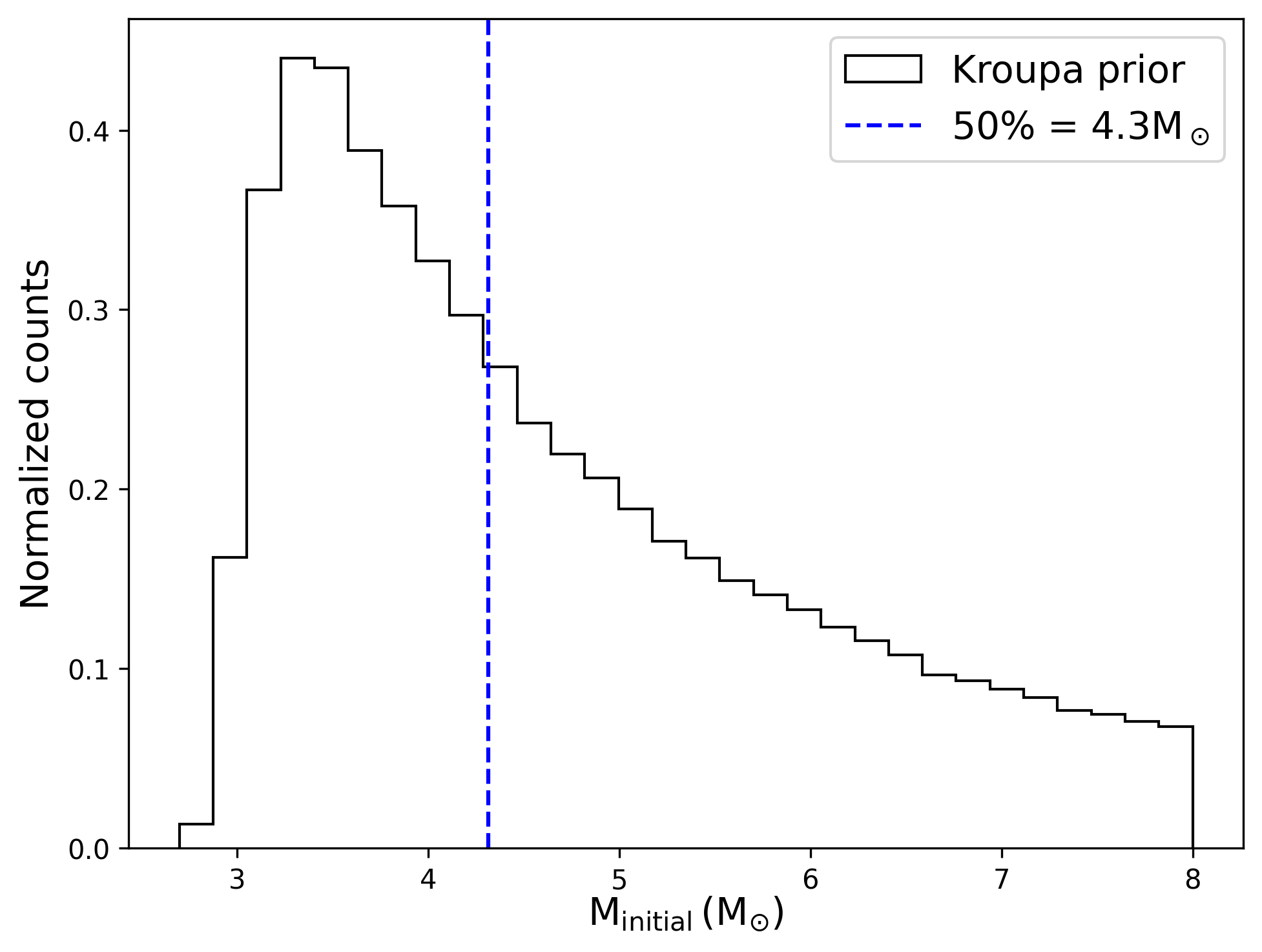}
    \caption{Progenitor mass distribution for a single candidate distribution based on the Kroupa IMF truncated sharply at $8M_\odot$, and gradually at $M_\text{turnoff}$ due to the uncertainty of the cluster age.}
    \label{fig:kroupa_prior}
\end{figure}

Now with only one independent variable $T_{\text{eff},2}$, we can solve for the approximate companion temperature with a simple grid search. This value is then used as an initial condition in the full binary fit. 

\subsection{Binary Fit} \label{subsec:binary}
The final step in obtaining a robust estimate of $T_{\text{eff},2}$ in systems with UV excess is a full SED fit to a combined MS+WD model. We employ the \textsc{Python} package \textsc{emcee} \citep{emcee} for this purpose, as it provides a comprehensive characterization of the posterior distribution, accounting for error bars on initial parameter estimates and revealing correlations between different parameters. Of the eight parameters required to construct the SED model, five are determined from previous analyses: $M_1,~M_2,~A_V,~[\text{Fe}/\text{H}]$ and the parallax $\varpi$. The remaining three parameters, $T_{\text{eff},1},~T_{\text{eff},2},~R_1$, are given uninformative priors. The walkers are initialized near the previously estimated values found in \autoref{subsec:uv_excess_appendix}, \autoref{subsec:optical_ir_appendix}. We run the \textsc{emcee} sampler with 30 walkers and 10,000 steps. The autocorrelation length of the final chain is approximately 100 steps. We discard the first 500 samples as burn-in and apply a thinning factor of 75 to the remaining samples. To address degeneracies that might lead the walkers to converge on solutions inconsistent with well-established parameters (such as parallax, masses, and metallicity), we employed truncated Gaussian priors-- see \autoref{tab:mcmc_priors}. For the uniform priors on $T_{\text{eff},1}$ and $R_1$ we selected broad limits that contain the expected range for a MS star. Ideally, for the uniform prior on $T_{\text{eff},2}$ we aim to use the broadest possible limits based on model constraints. However, in some cases, we had to impose a physical lower bound on the temperature to achieve convergence. This lower bound on $T_{\text{eff},2}$ is determined by the age of the host cluster, which sets the maximum possible cooling age for the WD. Through the cooling track, this translates into a minimum allowable temperature.
\begin{deluxetable}{lc}
\tablecaption{Different parameters and their prior distributions used for the MCMC fit. In some cases, we enforce a lower bound on $T_{\text{eff},2}$ using the minimal cooling temperature implied by the age of the cluster. Truncated Gaussians were used to prevent convergence on degenerate but nonphysical solutions.\label{tab:mcmc_priors}}
\tablehead{\colhead{Parameter} & \colhead{Prior}}
\startdata
$T_{\text{eff},1}$ & Uniform $3500-15000$\,K \\  
$T_{\text{eff},2}$ & Uniform on model range \\  
$R_1$ & Uniform $0.1-5R_\odot$ \\  
$M_1$ & Gaussian truncated at $3\sigma$ \\  
$M_2$ & Gaussian truncated at $3\sigma$ \\  
$A_V$ & Gaussian truncated at 0 \\  
$\varpi$ & Gaussian truncated at $3\sigma$ \\  
$[\text{Fe}/\text{H}]$ & Gaussian truncated at $3\sigma$ 
\enddata
\end{deluxetable}

\begin{deluxetable*}{cccccccccc} 
\tablecaption{SED Fitting Results. Both $\chi^2_\text{IR}$ and $\chi^2_\text{opt}$ are calculated using both optical and IR data, and normalized by the number of degrees of freedom ($N_\text{IR}+N_\text{opt} - 2$). However, $\chi^2_\text{opt}$ is calculated using the model fit to optical data only. Candidates 101, 140, 160, 192, 194, 282 are suspected of showing IR excess, and therefore are not given a full optical+IR fit. $T_\text{eff,2}$ values are obtained from the full binary fit for candidates with UV excess. This table is available online at \url{https://doi.org/10.5281/zenodo.14745560}. \label{tab:sed_results}} 
\tabletypesize{\small}
\tablewidth{0pt}
\tablecolumns{10}
\tablehead{
\colhead{Candidate \#} & \colhead{N$_\text{opt}$} & \colhead{N$_\text{IR}$} & \colhead{$\chi^2_\text{opt}$} & \colhead{$\chi^2_\text{IR}$} & \colhead{$T_\text{eff,1}$}& \colhead{$R_1$} & \colhead{$T_\text{eff,2}$} & \colhead{NUV Excess} & \colhead{FUV Excess} \\
\colhead{} & \colhead{\scriptsize Data points in optical}& \colhead{\scriptsize Data points in IR} & 
\colhead{\scriptsize Optical only fit} & 
\colhead{\scriptsize Optical+IR fit} & 
\colhead{\scriptsize (K)} & 
\colhead{\scriptsize ($R_\odot$)} & 
\colhead{\scriptsize (K)} &
\colhead{\scriptsize (erg cm$^{-2}$ s$^{-1}$ $\AA^{-1}$)}& 
\colhead{\scriptsize (erg cm$^{-2}$ s$^{-1}$ $\AA^{-1}$)}
}\startdata
6 & 8 & 6 & 0.79 & 0.68 & $8200 \pm 200$ & $1.99 \pm 0.06$ & \nodata & \nodata & \nodata \\ 
16 & 8 & 6 & 1.03 & 0.85 & $8100 \pm 200$ & $1.86 \pm 0.05$ & \nodata & \nodata & \nodata \\ 
22 & 7 & 6 & 0.63 & 0.26 & $4710 \pm 60$ & $0.8 \pm 0.02$ & \nodata & \nodata & \nodata \\ 
33 & 8 & 6 & 1.94 & 1.53 & $10200 \pm 400$ & $2.72 \pm 0.09$ & \nodata & \nodata & \nodata \\ 
48 & 8 & 6 & 0.53 & 0.52 & $6250 \pm 90$ & $1.49 \pm 0.04$ & \nodata & \nodata & \nodata \\ 
53 & 3 & 6 & 0.46 & 0.06 & $5900 \pm 100$ & $1.29 \pm 0.04$ & $11100 \pm 100$ & \nodata & 1.58e-16 \\ 
68 & 8 & 6 & 4.38 & 4.09 & $7000 \pm 100$ & $1.7 \pm 0.05$ & \nodata & \nodata & \nodata \\ 
71 & 8 & 6 & 0.63 & 0.63 & $6290 \pm 90$ & $1.3 \pm 0.04$ & \nodata & \nodata & \nodata \\ 
80 & 8 & 5 & 0.49 & 0.48 & $6500 \pm 100$ & $1.98 \pm 0.06$ & \nodata & \nodata & \nodata \\ 
96 & 8 & 6 & 1.83 & 1.27 & $6200 \pm 100$ & $1.42 \pm 0.04$ & \nodata & \nodata & \nodata \\ 
101 & 8 & 5 & 2.07 & \nodata & $5410 \pm 70$ & $1.05 \pm 0.03$ & \nodata & \nodata & \nodata \\ 
105 & 8 & 6 & 0.81 & 0.79 & $6800 \pm 100$ & $1.55 \pm 0.04$ & \nodata & \nodata & \nodata \\ 
110 & 8 & 6 & 1.77 & 1.72 & $7700 \pm 200$ & $2.39 \pm 0.07$ & \nodata & \nodata & \nodata \\ 
114 & 8 & 6 & 0.86 & 0.86 & $8200 \pm 200$ & $1.68 \pm 0.05$ & \nodata & \nodata & \nodata \\ 
121 & 8 & 5 & 1.66 & 1.08 & $10100 \pm 400$ & $1.84 \pm 0.06$ & \nodata & \nodata & \nodata \\ 
133 & 8 & 6 & 4.98 & 3.79 & $5910 \pm 90$ & $2.4 \pm 0.07$ & $22000 \pm 1000$ & \nodata & 4.31e-16 \\ 
140 & 8 & 6 & 2.83 & \nodata & $4460 \pm 50$ & $1.91 \pm 0.05$ & \nodata & \nodata & \nodata \\ 
154 & 8 & 6 & 1.68 & 1.16 & $6600 \pm 100$ & $1.87 \pm 0.05$ & \nodata & \nodata & \nodata \\ 
160 & 8 & 5 & 1.73 & \nodata & $4770 \pm 40$ & $0.85 \pm 0.02$ & \nodata & \nodata & \nodata \\ 
174 & 8 & 6 & 0.69 & 0.51 & $5290 \pm 50$ & $0.86 \pm 0.02$ & $8800 \pm 800$ & 5.16e-17 & \nodata \\ 
175 & 3 & 6 & 0.46 & 0.21 & $6300 \pm 100$ & $1.42 \pm 0.04$ & $11500 \pm 200$ & 4.04e-15 & 5.96e-16 \\ 
178 & 8 & 6 & 2.4 & 1.42 & $6200 \pm 100$ & $1.54 \pm 0.04$ & \nodata & \nodata & \nodata \\ 
180 & 8 & 5 & 1.74 & 0.83 & $6200 \pm 100$ & $1.84 \pm 0.05$ & \nodata & 5.46e-16 & \nodata \\ 
183 & 8 & 6 & 2.67 & 2.2 & $6500 \pm 100$ & $1.78 \pm 0.05$ & \nodata & \nodata & \nodata \\ 
191 & 7 & 6 & 1.17 & 0.35 & $4270 \pm 40$ & $0.78 \pm 0.02$ & \nodata & \nodata & \nodata \\ 
192 & 8 & 6 & 1.35 & \nodata & $5090 \pm 50$ & $0.84 \pm 0.02$ & \nodata & \nodata & \nodata \\ 
194 & 8 & 4 & 8.58 & \nodata & $4670 \pm 60$ & $1.44 \pm 0.04$ & \nodata & \nodata & \nodata \\ 
233 & 7 & 6 & 0.55 & 0.53 & $4610 \pm 60$ & $0.99 \pm 0.03$ & \nodata & \nodata & \nodata \\ 
236 & 8 & 6 & 0.56 & 0.43 & $6160 \pm 90$ & $1.29 \pm 0.04$ & $20000 \pm 4000$ & 1.73e-16 & \nodata \\ 
246 & 7 & 6 & 1.03 & 0.3 & $4160 \pm 50$ & $0.75 \pm 0.02$ & \nodata & \nodata & \nodata \\ 
248 & 6 & 5 & 0.47 & 0.14 & $3810 \pm 40$ & $0.74 \pm 0.02$ & \nodata & \nodata & \nodata \\ 
249 & 8 & 6 & 0.52 & 0.47 & $7000 \pm 100$ & $1.56 \pm 0.04$ & $17600 \pm 900$ & \nodata & 8.62e-16 \\ 
260 & 8 & 3 & 0.63 & 0.63 & $7400 \pm 200$ & $1.93 \pm 0.07$ & \nodata & \nodata & \nodata \\ 
261 & 8 & 3 & 0.71 & 0.49 & $4930 \pm 70$ & $0.87 \pm 0.03$ & \nodata & \nodata & \nodata \\ 
262 & 8 & 3 & 0.92 & 0.82 & $6400 \pm 100$ & $2.44 \pm 0.09$ & \nodata & \nodata & \nodata \\ 
270 & 8 & 5 & 1.11 & 1.1 & $8100 \pm 200$ & $2.13 \pm 0.06$ & \nodata & \nodata & \nodata \\ 
277 & 8 & 5 & 0.76 & 0.76 & $7900 \pm 200$ & $1.94 \pm 0.06$ & \nodata & \nodata & \nodata \\ 
281 & 8 & 6 & 0.92 & 0.65 & $4610 \pm 40$ & $0.74 \pm 0.02$ & $7300 \pm 100$ & 5.35e-17 & \nodata \\ 
282 & 8 & 6 & 4.29 & \nodata & $4570 \pm 40$ & $0.82 \pm 0.02$ & \nodata & \nodata & \nodata \\ 
283 & 7 & 5 & 0.51 & 0.46 & $4630 \pm 60$ & $0.84 \pm 0.02$ & $9100 \pm 600$ & 2.69e-17 & \nodata \\ 
\enddata
\end{deluxetable*}

\section{IFMR} \label{appendix:ifmr}
\subsection{Kroupa Prior}
In \autoref{sec:ifmr}, we apply simple considerations to constrain the progenitor masses of our systems, even in the absence of direct measurements of the WD companion. The progenitor mass must lie between the cluster's MS turnoff mass and the WD formation limit of $8\,\text{M}_\odot$. Additionally, since lower-mass stars are more common than higher-mass stars, we assume a Kroupa initial mass function (IMF; power-law index $\alpha = 2.3$, \citealt{kroupa_imf}) to model the distribution of progenitor masses. This prior distribution, referred to as the `Kroupa prior', is illustrated in \autoref{fig:kroupa_prior}.

To estimate $M_\text{initial}$, we draw 100 samples from the Kroupa IMF within the mass range $M_\text{turnoff}$ to $8\,\text{M}_\odot$. To account for uncertainty in the cluster age, which affects $M_\text{turnoff}$, we repeat this process 1,000 times, sampling the cluster age from a normal distribution in each iteration. These repeated samples are combined to refine the $M_\text{initial}$ distribution. The median of the resulting distribution is adopted as the best estimate of $M_\text{initial}$, while the 16th and 84th percentiles define the lower and upper bounds of the uncertainty.

In \autoref{fig:ifmr_groups}, we present the IFMR of the full sample, derived from all three calculation methods of $M_\text{initial}$: `no UV,' `no excess,' and `UV excess'. The more reliable `UV excess' group is highlighted with encircled markers.

\subsection{UV Excess}
In \autoref{subsec:ifmr_uv_excess}, we use WD temperatures for systems with UV excess to calculate the progenitor mass. Each candidate has a posterior distribution of $T_{\text{eff},2}$ obtained from the MCMC binary SED fit (\autoref{subsec:sed_analysis}). We resample this posterior by drawing $5,000$ instances of $T_{\text{eff},2}$. Additionally, we draw $5,000$ instances each of the cluster age $\tau_\text{tot}$, the WD mass $M_2$, and the metallicity from Gaussian distributions. For each set of these values, we calculate the WD cooling age $\tau_\text{cool}$ using the cooling tracks in \autoref{tab:massradius}. Subtracting $\tau_\text{cool}$ from $\tau_\text{tot}$ gives the progenitor lifetime $\tau_\text{life}$, from which we derive $5,000$ instances of the progenitor mass $M_\text{initial}$. This is the `unweighted' $M_\text{initial}$ distribution shown in \autoref{fig:m_initial_posterior}.

We retain $5,000$ instances with $M_\text{turnoff}<M_\text{initial}<8\,\text{M}_\odot$, as masses outside this range would not contribute once weights are applied. Finally, we apply weights based on the Kroupa prior, suppressing higher progenitor masses and enforcing the $M_\text{turnoff}$ and $8\,\text{M}_\odot$ limits. The resulting distribution is the `weighted' distribution shown in \autoref{fig:m_initial_posterior}.

\subsection{No UV Excess}
In \autoref{subsec:ifmr_no_excess}, we use the maximal WD temperature $T_\text{hottest}$, derived from the absence of UV excess in the SED, to refine the Kroupa prior for systems with no UV excess. We generate $1000$ instances of the cluster age $\tau_\text{tot}$ and the WD mass $M_2$, compute the WD cooling age (corresponding to $T_\text{hottest}$), and subtract this from $\tau_\text{tot}$ to obtain $\tau_\text{life}$. For each instance, we calculate the corresponding initial mass $M_\text{hottest}$. Since a cooler WD implies a longer cooling time and thus a more massive progenitor, $M_\text{hottest}$ serves as a lower limit for the Kroupa prior.

\begin{figure*}
    \centering
    \includegraphics[width=0.8\textwidth]{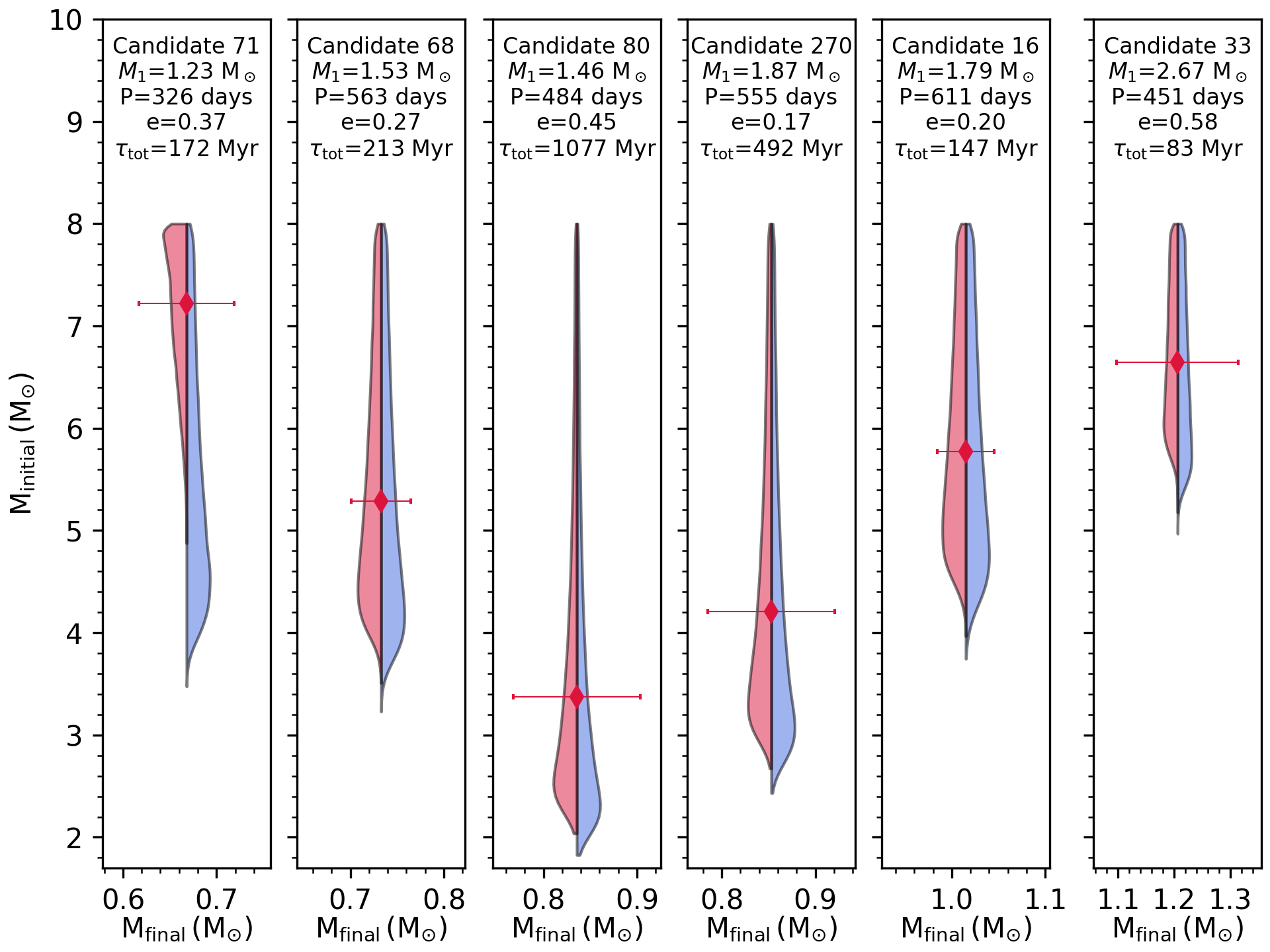}
    \caption{Tentative IFMR for the six systems with UV photometry, but no UV excess. The right side of the violin (blue) visualizes the probability distribution of $M_\text{initial}$ according to the Kroupa-based prior, while the left side (red) also includes the $M_\text{hottest}$ constraint from the SED (where available).  Only in candidate 71 the uncertainty on $M_{\text{initial}}$ was narrowed down. For the rest, a luminous primary can hide almost any WD in the $NUV$ band---so we were not able to put a tighter constraint on the progenitor mass.}
    \label{fig:ifmr_no_excess}
\end{figure*}

We then draw $100$ masses from the Kroupa IMF, truncated at $M_\text{hottest}$ and $8\,\text{M}_\odot$, for each of the $1000$ iterations. This produces a distribution of $M_\text{initial}$ with $100,000$ instances, incorporating both the WD temperature constraint and uncertainties in $\tau_\text{tot}$ and $\tau_\text{cool}$. A comparison of the refined $M_\text{initial}$ distributions with the Kroupa-based prior is shown in \autoref{fig:ifmr_no_excess}.

\begin{deluxetable*}{ccccccccccccc} 
\tablecaption{Sample data for our final 33 MS+WD candidates. Additionally, we provide data for one MS+SD candidate (180). This table is available online at \url{https://doi.org/10.5281/zenodo.14742239}.\label{tab:candidate_table}} 
\tabletypesize{\scriptsize}
\tablewidth{0pt}
\tablecolumns{13}
\tablehead{
\colhead{Candidate \#} & \colhead{Gaia DR3 ID} & \colhead{Cluster} & \colhead{$M_1$} & \colhead{$M_\text{final}$} 
& \colhead{$M_\text{initial}$} & \colhead{Age} & \colhead{[Fe/H]} & \colhead{$A_v$} & \colhead{$FUV$} & \colhead{$NUV$} \\
\colhead{} & \colhead{} & \colhead{} &
\colhead{($\text{M}_\odot$)} & \colhead{($\text{M}_\odot$)} & \colhead{($\text{M}_\odot$)} & \colhead{(Myr)} & \colhead{(dex)} & \colhead{(mag)} & \colhead{(mag)} & \colhead{(mag)}
}
\startdata
6 & 183325907326150528 & NGC 1912 & $1.85 \pm 0.03$ & $0.94 \pm 0.12$ & $4.32^{+1.86}_{-0.94}$ & $363 \pm 49$ & $0.05 \pm 0.03$ & $0.7 \pm 0.01$ & \nodata & \nodata \\ 
16 & 422751230068950016 & HSC 924 & $1.79 \pm 0.02$ & $1.02 \pm 0.03$ & $5.77^{+1.33}_{-0.9}$ & $148 \pm 19$ & $-0.02 \pm 0.11$ & $0.71 \pm 0.03$ & \nodata & $15.93 \pm 0.01$ \\ 
22 & 457836680469874560 & CWNU 1076 & $0.77 \pm 0.01$ & $0.39 \pm 0.02$ & $6.57^{+0.9}_{-0.76}$ & $80 \pm 11$ & $-0.08 \pm 0.06$ & $0.49 \pm 0.04$ & \nodata & \nodata \\ 
33 & 507246430918975360 & HSC 1090 & $2.67 \pm 0.06$ & $1.21 \pm 0.11$ & $6.64^{+0.86}_{-0.66}$ & $84 \pm 5$ & $0.04 \pm 0.16$ & $1.85 \pm 0.05$ & \nodata & $17.56 \pm 0.05$ \\ 
48 & 534740028404333184 & Collinder 463 & $1.31 \pm 0.02$ & $0.62 \pm 0.04$ & \nodata & $58 \pm 8$ & $0.05 \pm 0.02$ & $1.01 \pm 0.01$ & \nodata & $20.32 \pm 0.12$ \\ 
53 & 661148268907314432 & NGC 2632 & $1.2 \pm 0.01$ & $0.8 \pm 0.01$ & $5.33^{+1.31}_{-0.89}$ & $761 \pm 97$ & $0.08 \pm 0.04$ & $0.01 \pm 0.01$ & $21.06 \pm 0.08$ & \nodata \\ 
68 & 1872574993400437888 & Theia 3916 & $1.53 \pm 0.02$ & $0.73 \pm 0.03$ & $5.29^{+1.53}_{-0.94}$ & $213 \pm 27$ & $0.13 \pm 0.03$ & $0.88 \pm 0.06$ & \nodata & $18.39 \pm 0.04$ \\ 
71 & 1942497610724886912 & Theia 401 & $1.23 \pm 0.02$ & $0.67 \pm 0.05$ & $7.22^{+0.59}_{-0.83}$ & $173 \pm 22$ & $-0.08 \pm 0.08$ & $0.29 \pm 0.03$ & \nodata & $18.25 \pm 0.03$ \\ 
80 & 1989228126220910208 & UBC 6 & $1.46 \pm 0.03$ & $0.84 \pm 0.07$ & $3.37^{+2.02}_{-0.86}$ & $1078 \pm 137$ & $-0.03 \pm 0.06$ & $0.39 \pm 0.01$ & \nodata & $18.50 \pm 0.08$ \\ 
96 & 2059201419867861248 & Theia 470 & $1.27 \pm 0.01$ & $0.63 \pm 0.02$ & $5.33^{+1.53}_{-0.94}$ & $172 \pm 22$ & $-0.01 \pm 0.06$ & $0.34 \pm 0.01$ & \nodata & \nodata \\ 
105 & 2073689276577590912 & CWNU 1183 & $1.45 \pm 0.01$ & $0.79 \pm 0.02$ & $7.58^{+0.31}_{-0.49}$ & $48 \pm 6$ & $0.02 \pm 0.04$ & $0.19 \pm 0.02$ & \nodata & \nodata \\ 
110 & 2166849556774987520 & NGC 6991 & $1.84 \pm 0.02$ & $0.99 \pm 0.09$ & $3.33^{+2.04}_{-0.85}$ & $897 \pm 55$ & $0.15 \pm 0.02$ & $0.5 \pm 0.02$ & \nodata & \nodata \\ 
114 & 2174275039828118912 & ASCC 114 & $1.68 \pm 0.02$ & $0.75 \pm 0.03$ & \nodata & $28 \pm 3$ & $-0.09 \pm 0.07$ & $1.06 \pm 0.02$ & \nodata & $17.49 \pm 0.04$ \\ 
121 & 2205542917130608512 & FSR 0398 & $2.02 \pm 0.03$ & $0.98 \pm 0.05$ & \nodata & $18 \pm 2$ & $0.07 \pm 0.1$ & $1.65 \pm 0.04$ & \nodata & $17.43 \pm 0.04$ \\ 
133 & 2919695270059319808 & vdBergh 83 & $1.54 \pm 0.04$ & $0.88 \pm 0.04$ & \nodata & $19 \pm 5$ & $-0.03 \pm 0.13$ & $0.07 \pm 0.0$ & $20.06 \pm 0.19$ & \nodata \\ 
154 & 3451094801741563264 & NGC 2099 & $1.49 \pm 0.02$ & $1.07 \pm 0.13$ & $3.67^{+2.01}_{-0.89}$ & $652 \pm 86$ & $0.14 \pm 0.02$ & $0.54 \pm 0.01$ & \nodata & \nodata \\ 
174 & 4782447359703077248 & HSC 2047 & $0.91 \pm 0.01$ & $0.45 \pm 0.01$ & \nodata & $318 \pm 43$ & $0.02 \pm 0.04$ & $0.02 \pm 0.01$ & \nodata & $19.24 \pm 0.09$ \\ 
175 & 4816334892186732160 & HSC 1986 & $1.27 \pm 0.03$ & $0.6 \pm 0.02$ & $2.81^{+0.24}_{-0.2}$ & $994 \pm 122$ & $-0.02 \pm 0.12$ & $0.01 \pm 0.01$ & $19.42 \pm 0.13$ & $13.95 \pm 0.01$ \\ 
178 & 5222629723121939328 & ASCC 51 & $1.32 \pm 0.01$ & $0.61 \pm 0.01$ & $5.56^{+1.42}_{-0.92}$ & $147 \pm 19$ & $-0.02 \pm 0.03$ & $0.18 \pm 0.01$ & \nodata & \nodata \\ 
180 & 5237947539322507136 & Theia 2745 & $1.44 \pm 0.02$ & $1.17 \pm 0.04$ & \nodata & $270 \pm 34$ & $0.12 \pm 0.06$ & $0.47 \pm 0.01$ & \nodata & $18.15 \pm 0.07$ \\ 
183 & 5242761476090953728 & UBC 511 & $1.45 \pm 0.02$ & $0.97 \pm 0.04$ & $3.43^{+2.04}_{-0.86}$ & $809 \pm 109$ & $0.06 \pm 0.06$ & $0.25 \pm 0.01$ & \nodata & \nodata \\ 
191 & 5263315540296196992 & HSC 2407 & $0.71 \pm 0.01$ & $0.32 \pm 0.01$ & $4.03^{+1.96}_{-0.93}$ & $456 \pm 61$ & $-0.01 \pm 0.08$ & $0.01 \pm 0.01$ & \nodata & \nodata \\ 
233 & 5541167076447007872 & UPK 502 & $0.84 \pm 0.01$ & $0.44 \pm 0.04$ & $5.74^{+1.34}_{-0.92}$ & $133 \pm 21$ & $-0.01 \pm 0.06$ & $0.27 \pm 0.01$ & \nodata & \nodata \\ 
236 & 5581250082066100736 & HSC 1961 & $1.2 \pm 0.01$ & $0.6 \pm 0.03$ & $5.58^{+0.96}_{-0.5}$ & $122 \pm 15$ & $-0.08 \pm 0.05$ & $0.11 \pm 0.01$ & \nodata & $17.57 \pm 0.04$ \\ 
246 & 5635800835352566144 & OCSN 86 & $0.7 \pm 0.01$ & $0.34 \pm 0.01$ & $4.01^{+1.93}_{-0.92}$ & $462 \pm 61$ & $0.01 \pm 0.04$ & $0.04 \pm 0.01$ & \nodata & \nodata \\ 
248 & 5752835434289715584 & AT 4 & $0.64 \pm 0.01$ & $0.39 \pm 0.04$ & $3.49^{+2.02}_{-0.87}$ & $740 \pm 97$ & $0.03 \pm 0.05$ & $0.07 \pm 0.01$ & \nodata & \nodata \\ 
249 & 5753128007461925760 & AT 4 & $1.45 \pm 0.01$ & $0.72 \pm 0.02$ & $2.83^{+0.21}_{-0.15}$ & $740 \pm 97$ & $0.03 \pm 0.05$ & $0.07 \pm 0.01$ & $18.70 \pm 0.10$ & $14.82 \pm 0.01$ \\ 
260 & 5870278497578560128 & Ruprecht 108 & $1.7 \pm 0.03$ & $1.19 \pm 0.07$ & $4.74^{+1.75}_{-0.96}$ & $263 \pm 34$ & $0.03 \pm 0.04$ & $0.47 \pm 0.01$ & \nodata & \nodata \\ 
261 & 5870538463339228416 & CWNU 1138 & $0.85 \pm 0.02$ & $0.44 \pm 0.02$ & $5.11^{+1.61}_{-0.95}$ & $197 \pm 24$ & $-0.02 \pm 0.1$ & $0.32 \pm 0.03$ & \nodata & \nodata \\ 
262 & 5930051213822024704 & NGC 6208 & $1.56 \pm 0.03$ & $1.17 \pm 0.14$ & $2.65^{+2.02}_{-0.73}$ & $1941 \pm 243$ & $0.11 \pm 0.04$ & $0.54 \pm 0.01$ & \nodata & \nodata \\ 
270 & 5990577714009072896 & Theia 1907 & $1.87 \pm 0.03$ & $0.85 \pm 0.07$ & $4.21^{+1.92}_{-0.94}$ & $493 \pm 65$ & $0.08 \pm 0.04$ & $1.0 \pm 0.01$ & \nodata & $17.31 \pm 0.04$ \\ 
277 & 6071002793352423424 & Stock 15 & $1.81 \pm 0.02$ & $0.96 \pm 0.09$ & $4.34^{+1.87}_{-0.95}$ & $364 \pm 49$ & $0.18 \pm 0.04$ & $0.72 \pm 0.01$ & \nodata & \nodata \\ 
281 & 6460313579841489920 & HSC 2718 & $0.77 \pm 0.01$ & $0.36 \pm 0.01$ & \nodata & $241 \pm 31$ & $0.07 \pm 0.08$ & $0.01 \pm 0.01$ & \nodata & $21.10 \pm 0.17$ \\ 
283 & 6909825338076583424 & UPK 51 & $0.79 \pm 0.01$ & $0.41 \pm 0.02$ & \nodata & $167 \pm 22$ & $-0.02 \pm 0.06$ & $0.15 \pm 0.01$ & \nodata & $22.03 \pm 0.33$ \\ 
\enddata
\end{deluxetable*}

\section{Excluded Systems and Analysis Inconsistencies}\label{appendix:excluded_systems}
In \autoref{fig:ifmr_groups}, we present only 26 data points, even though 34 systems met all the criteria outlined in \autoref{sec:analysis}. We turn to discuss the inconsistencies identified during the analysis of the remaining 8 systems as well as providing potential explanations. Investigating these inconsistencies is crucial because they reveal potential weaknesses in our method, allowing us to refine and improve it. Moreover, some of the issues identified may also be present in the `no UV' group within our sample, even if they have not yet been detected. Four candidates showed problems at the SED analysis stage (\autoref{subsec:sed_analysis}): candidates 174, 180, 281, and 283. All four belong to the UV excess group and have a successful binary fit. The remaining systems were dropped during the derivation of $M_\text{initial}$ (\autoref{subsec:ifmr_no_excess}): candidates 48, 114, 121 and 133. All four of these have UV data but only 133 has UV excess and a successful binary fit.

\emph{Candidate 180}: The SED failed to fit a MS+WD model. However, a MS+blackbody model can successfully account for the $NUV$ excess of this system if $R_2 \approx 0.15\,\text{R}_\odot$ and $T_{\text{eff},2} \approx 12{,}000\,\text{K}$. This suggests that the companion in this system might be a hot subdwarf (SD). Due to its small radius and high temperature, the companion would be much fainter than the primary star in visible wavelengths.  Nevertheless, this conclusion is not definitive because the dynamic mass $M_2 = 1.16\,\text{M}_\odot$ is too high for a typical SD \citep{heber_hot_sdb}. A faint massive tertiary—possibly a very close companion to the SD—might be required to fully explain the system. This makes candidate 180 an interesting target for follow-up studies, as SD stars are thought to be products of mass transfer and represent an intermediate phase on the evolutionary path to becoming a WD.

\emph{Candidate 174}: The SED MS+WD fit suggests that this system hosts a $M_2=0.45\text{M}_\odot$, $T_{\text{eff},2}= 8750\,\pm 800$K WD companion. The cluster age is $317\pm 40\,\text{Myr}$ in this system, at first glance too young to allow a WD of this mass to cool to $8750\,$K. A more detailed analysis shows this fast cooling is possible-- but only on the fringes of the $\tau_\text{tot},T_{\text{eff},2}$ distributions, with a probability of about $\sim 1/5000$. This is significant enough to raise a red flag, and we believe the mismatch between the WD cooling age and the cluster age is most likely due to a false cluster association. While Gaia DR3 single-object parallaxes and proper motions for candidate 174, as identified by \citet{Hunt_2023}, match the cluster distribution, the more accurate NSS values place it outside the cluster. This suggests it is a foreground object misidentified as a member due to astrometric motion from its binary nature.

\emph{Candidate 48}: This system with a $M_2 = 0.62\text{M}_\odot$ companion belongs to a young open cluster $\tau_\text{tot} = 57\pm 7\,\text{Myr}$. A theoretical cooling track suggests that a WD in this cluster must have $T_{\text{eff},2} > 21{,}000\pm 1000$K. However, the analysis done on `no excess' systems in \autoref{subsec:sed_analysis} suggests any WD companion with $T_{\text{eff},2} > 19{,}500\,$K should have produced a noticeable $NUV$ excess, which is not observed in the SED. Unlike candidate 174, the cluster association of 48 seems reliable. We are left to conclude that the most likely source of error lies in an underestimation of the cluster age uncertainty (\autoref{subsec:cluster_analysis}). Accurately determining the ages of young open clusters under $100\,\text{Myr}$ is inherently challenging: the MS turnoff is poorly sampled due to the rarity of massive stars, and the best-fitting cluster age is highly sensitive to typically just one star.  Even our fitting algorithm, designed to address such issues, may not fully capture this uncertainty in its calculations. Moreover, neglecting the effects of stellar rotation can lead to a systematic underestimation of young cluster ages, as discussed by \citet{Cummings_clusters_2018}. More generous error bars would likely provide a better representation of our knowledge of the cluster's age, and help resolve the apparent tension between the cooling age and cluster age for this system.

\emph{Candidates 114, 121, 133}: These systems are members of open clusters with ages of $28 \pm 3$, $18 \pm 2$, and $19 \pm 5\,\text{Myr}$, respectively, and have companion masses of $M_2 = 0.75$, $0.98$, and $0.88\,\text{M}_\odot$, respectively. In all three, the cluster association seems secure. Candidates 114 and 121 have UV but no excess, while candidate 133 has a successful MS+WD binary fit. These clusters appear too young to allow an MS star with $M_\text{initial} < 8\,\text{M}_\odot$ to evolve into a WD. As discussed for candidate 48, determining the ages of young clusters is inherently challenging, and the uncertainties may be significantly underestimated. Larger error bars could reconcile this apparent inconsistency. For candidate 121, an alternative explanation is that the companion is a neutron star, which would address both the cluster age/WD formation time tension and the absence of UV excess.

\emph{Candidates 281, 283}: The SEDs of both systems have been successfully fit with a MS+WD model, yielding best-fit values of $T_{\text{eff},2} = 7300\pm 150\,\text{K}$ and $9000\pm 600\,\text{K}$, respectively. Their cluster associations appear reliable, with age estimates of $240 \pm 30\,\text{Myr}$ and $167 \pm 20\,\text{Myr}$, showing no apparent issues. However, cooling models rule out the possibility that helium-core WDs with $M_2 = 0.36$ and $0.41\,M_\odot$ could reach such low temperatures within these cluster ages. The discrepancy becomes more pronounced when considering the uncertainties in ages and temperatures, as no combination of these values resolves the disagreement. Specifically, candidate 281 is too cold by approximately $5,300$\,K,\footnote{This value refers to the $99_\text{th}$ percentile of the temperature distribution. The median is even colder than that!} requiring an additional $1.4$ Gyr of cooling time to reconcile the difference, while candidate 283 is about $4,000$\,K too cold, necessitating an extra $700$ Myr of cooling time. For these two systems, we were unable to find a satisfactory explanation for the cluster age versus cooling age inconsistency. This could indicate a flaw in our methods or calculations, or perhaps point to a real physical phenomenon that is not yet fully understood.

\section{Selection Bias}\label{appendix:bias}

For each open cluster, we define the selection probability of a WD with mass $M_2$ as $p(M_2)$. However, the Gaia NSS and AMRF triage selection probabilities depend on the system's orbital parameters and distance. Using the total probability formula:  
\begin{equation}\label{eq:total_prob}
p(M_2) = \sum_{P, M_1, \varpi} p(P, M_1, \varpi) \, p(M_2 \mid P, M_1, \varpi),
\end{equation}
we focus on parallax ($\varpi$), orbital period ($P$), and primary mass ($M_1$) as the main factors influencing selection.

Assuming a fixed parallax for each cluster, as average distance dominates detection capability, and also assuming independent distributions for $P$ and $M_1$ (i.e., $p(P, M_1) = p(P) p(M_1)$), we derive:
\begin{equation}\label{eq:total_prob2}
p(M_2) = \sum_{P, M_1} p(P) \, p(M_1) \, p(M_2 \mid P, M_1, \varpi).
\end{equation}
Here, $p(P)$ is chosen as the period distribution of Gaia astrometric binaries passing \citet{halbwachs2022gaia} cuts, while $p(M_1)$ is taken from each cluster’s observed mass distribution via PARSEC isochrones. Secondary factors (e.g., eccentricity, sky position) are omitted from \autoref{eq:total_prob2} for simplicity but are implicitly accounted for in the calculations throughout this subsection.

For each cluster, we calculate $p(M_2)$ for 5 values of $M_2$ between $0.4-1.2\,\text{M}_\odot$. We divide the period and primary mass range into 5 bins, using the code of \citet{nss_sf} to estimate the NSS detection probability for each $\left(M_2,M_1,P,\varpi\right)$ configuration. For each configuration, we also estimate the AMRF $\mathcal{A}$, the maximal AMRF for a binary MS $\mathcal{A}_{\textit{MS}}$ and the \texttt{class-I probability}, all described in \autoref{subsec:amrf_analysis}. The selection probability of each configuration is the NSS detection probability if it passes the AMRF cut (\texttt{class-I probability}$<10\%$), or 0 otherwise. Lastly, we sum over all configurations for each cluster according to \autoref{eq:total_prob2}. \autoref{fig:bias_vs_distance} illustrates the implications of the calculations described above.

\bibliography{ifmr_binary_wd}{}
\bibliographystyle{aasjournal}

\end{document}